\documentclass[journal]{new-aiaa}
\usepackage[utf8]{inputenc}
\usepackage{textcomp}
\usepackage{wrapfig}
\usepackage{url}
\usepackage{graphicx}
\usepackage{booktabs}
\usepackage{amsmath}
\usepackage[version=4]{mhchem}
\usepackage{siunitx}
\usepackage[noend]{algcompatible}
\usepackage{algorithm} 
\usepackage{longtable,tabularx}
\usepackage{threeparttable}
\algnewcommand\INPUT{\item[\textbf{Input:}]}%
\algnewcommand\OUTPUT{\item[\textbf{Output:}]}%

\title{Hierarchical Framework for Optimizing Wildfire Surveillance and Suppression using Human-Autonomous Teaming}

\author{Mahdi Al-Husseini, \footnote{MS Student, Aeronautics \& Astronautics Dept., Stanford University, Stanford, CA 94305, Student Member.}
Kyle H. Wray, \footnote{Visiting Scholar, Aeronautics \& Astronautics Dept., Stanford University, Stanford, CA 94305, Member.}
and Mykel J. Kochenderfer \footnote{Associate Professor, Aeronautics \& Astronautics Dept., Stanford University, Stanford, CA 94305, Associate Fellow.}}
\affil{Stanford Intelligent Systems Laboratory, Stanford, CA, 94305}

\begin{document}

\maketitle

\begin{abstract}
The integration of manned and unmanned aircraft can help improve wildfire response. Wildfire containment failures occur when resources available to first responders, who execute the initial stages of wildfire management referred to as the initial attack, are ineffective or insufficient. Initial attack surveillance and suppression models have linked action spaces and objectives, making their optimization computationally challenging. The initial attack may be formulated as a multi-agent partially observable Markov decision process (MPOMDP). We divide the initial attack MPOMDP into surveillance and suppression processes with their respective planners operating on different, but constant, time scales. A hierarchical framework iterates between surveillance and suppression planners while also providing collision avoidance. This framework is exemplified by a set of multi-rotor unmanned aircraft surveying an initial attack fire while a manned helicopter suppresses the fire with a water bucket. Wildfire-specific solver extensions are formulated to reduce the otherwise vast action spaces. The hierarchical framework outperforms firefighting techniques and a myopic baseline by up to 242\% for moderate wildfires and 60\% for rapid wildfires when simulated in abstracted and actual case studies. We also validate the early dispatching of additional suppression assets using regression models to ensure wildfire containment to thresholds established by wildfire agencies.
\end{abstract}

\section*{Nomenclature}
{\renewcommand\arraystretch{1.0}
\noindent\begin{longtable*}{@{}l @{\quad=\quad} l@{}}
$x$           & wildfire cell \\
$F(x)$        & fuel remaining in $x$ \\
$\beta(x)$    & fuel reduced in $x$ due to suppressive activities at $T$ \\
$\mathcal{B}(x)$        & whether $x$ is on fire in belief map $\mathcal{B}$ \\
$\mathcal{U}(x)$        & uncertainty regarding state of $x$ in belief map $\mathcal{B}$ \\
$R(x)$        & resources on $x$ at time $t$ = 0 \\
$\mathcal{D}(x)$        & instantaneous destruction by wildfire on $x$ in belief map $\mathcal{B}$ with respect to $R(x)$ \\
$\mathcal{W}(x)$       & whether $x$ is on fire in the actual wildfire map $\mathcal{W}$ \\
$t$, $d_t$           & time-step such that $t$ = 0 is the incidence of the wildfire, duration of $t$ \\
$k$           & function of distance between wildfire and suppression source and manned aircraft cruising speeds  \\
$T$, $d_T$           & time-step such that $T$ = 1 denotes the first act of wildfire suppression, duration of $T$ equal to $k * d_t$ \\
$\alpha$      & expended unit of fuel in cell $x$ in $t$  \\
$\delta(x)$   & probability that $x$ is suppressed in $T$  \\
$P(x)$        & probability that $x$ ignites in $t$  \\
$p(x, x')$    & probability that $x'$ ignites $x$ at $t+1$ \\
$F_T$,  $P_T$       & set of $x$ fully suppressed at $T$, set of $x$ partially suppressed at $T$ \\
$\gamma_F$, $\gamma_P$ & fuel reduced for a fully suppressed cell, fuel reduced for a partially suppressed cell \\
$V_{e}$       & set of $x$ surveyed given action $e$ \\
$R_o$         & surveillance reward for belief map updates \\
$P_u$         & penalty for unmanned aircraft proximity to one another \\
$D_u$         & distance between unmanned aircraft below which $P_u$ is incurred \\
$P_m$         & penalty for unmanned aircraft proximity to manned aircraft \\
$D_m$         & distance between unmanned aircraft and manned aircraft below which $P_m$ is incurred \\
$P_i$         & penalty for distance between unmanned aircraft and $IA_O$ \\
$(X_{u_1}, Y_{u_1}, Z_{u_1})$     & location in $x$, $y$, $z$ of first unmanned aircraft $U_1$ \\
$(X_{u_2}, Y_{u_2}, Z_{u_2})$     & location in $x$, $y$, $z$ of second unmanned aircraft $U_2$ \\
$\mathcal{S}$ & surveillance state composed of $(X_{u_1}, Y_{u_1}, Z_{u_1}, X_{u_2}, Y_{u_2}, Z_{u_2})$ \\
$e / E$           & surveillance action / set of surveillance actions \\
$\mathcal{H}$ & suppression state composed of ($X_m, Y_m, DR$) \\
$(X_{m}, Y_{m})$ & location in $x$, $y$  of water drop  \\
$DR$          & drop type \\
$2^{|P_T|}$   & all outcomes of the partially suppressed set ($2^{18}$) \\
$a / A$           & suppression action / set of suppression actions \\
$R_M$        & suppression reward for localized resource destruction minimization model \\
$P_M$        & suppression penalty for global resource destruction minimization model \\
$AOA$        & manned aircraft axis of advance \\
$IA_O$    & initial attack fire origin \\
$\mathcal{R}$       & historical wildfire ring array
\end{longtable*}}

\section{Introduction}
\lettrine{E}{ach} year since 2000, an average of 70,600 wildfires have burned through a cumulative seven million acres, resulting in tens of billions of dollars in damages and thousands of lives lost. Wildfires in 2020 alone were responsible for more than 17,000 destroyed structures and 3,500 fatalities \cite{crswildfire2021}. The initial attack occurs when a first set of dispatched assets responds to an incipient wildfire \cite{hirsch1998using}. Initial attack effectiveness significantly influences the outcome of a wildfire's eventual containment, placing notable burden on the part of first responders \cite{rahn2010initial}. The California Department of Forestry and Fire Protection (CALFIRE) and others define initial attack response success as maintaining 95\% of wildfires under 10 acres \cite{eldorado2022}. The integration of manned and unmanned aircraft as part of an initial attack is introduced to \textit{improve wildfire visibility} and \textit{minimize initial attack fire damage}, while further serving to \textit{ensure fire containment to 10 acres} through the early requisition of additional suppression assets. \\
\indent
Although there exists ample literature on the optimized maneuver of distributed unmanned aircraft to conduct tactical wildfire management, there is less focus on hierarchical coordination of surveillance and suppression operations, and even less research integrating both manned and unmanned aircraft. To our knowledge, this paper is the first to do all three: optimize initial attack surveillance and suppression activities using a hierarchical framework of asynchronous planners to support a human-autonomous aircraft team.
\begin{wrapfigure}{L}{0.55\textwidth}
\centering
\includegraphics[angle=270,width=0.55\textwidth]{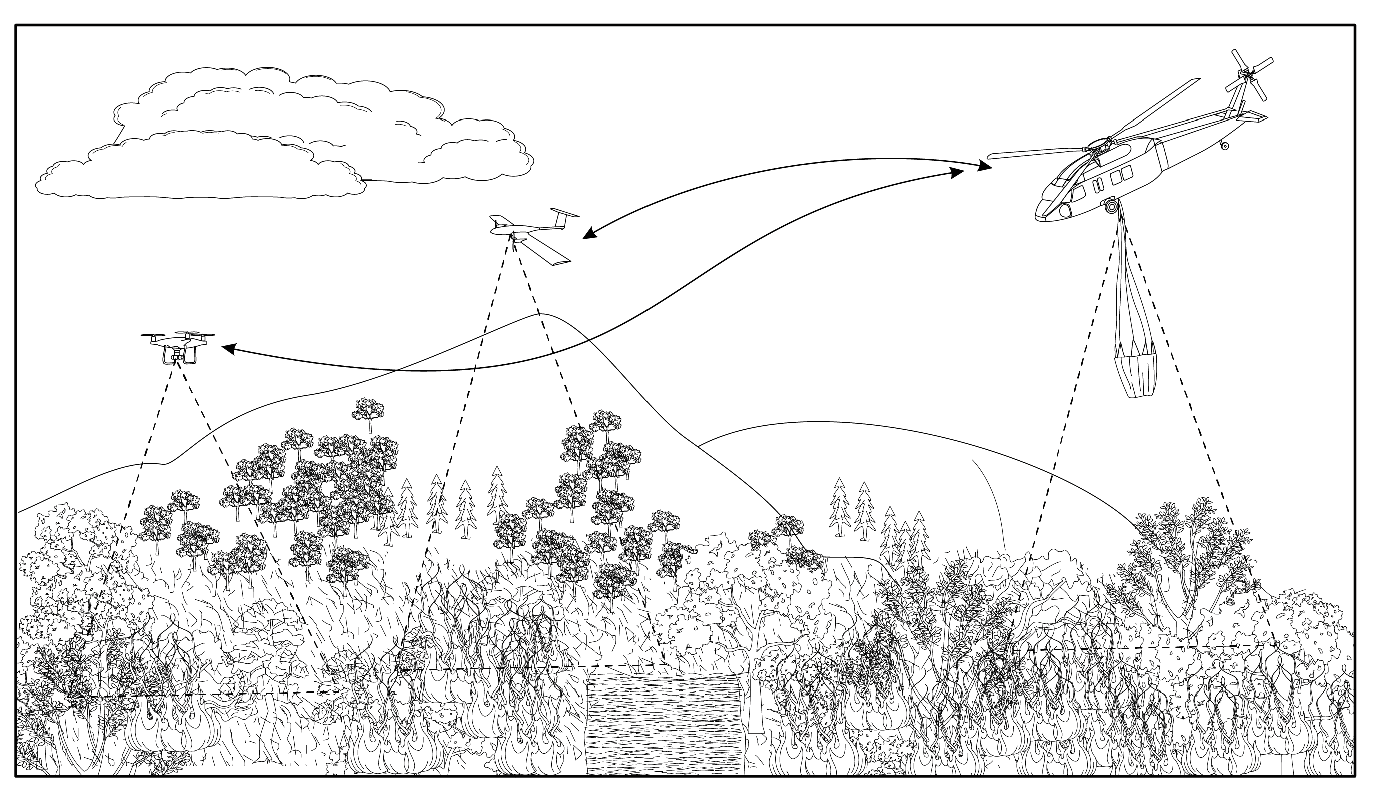}
\caption{Manned and unmanned aircraft coordinating to survey and suppress an initial attack fire.}
\end{wrapfigure}
Human-autonomous teaming involves manned and unmanned agents collaborating to optimize mission execution \cite{o2022human}. Teaming in aviation applications frees flight crews to focus on complex mission-essential tasks, such as wildfire suppression, while allowing critical but routine secondary tasks, such as wildfire surveillance, to be automated \cite{gaydos2014manned}. Despite the opportunity presented by unmanned aircraft, manned aircraft are expected to play a critical role in wildfire response into the foreseeable future. CALFIRE, the premier firefighting aviation program in the world with more firefighting aircraft than any other, continues to rely almost exclusively on manned aircraft for surveillance and suppression \cite{calfire2019}. Integrating low and medium-altitude  unmanned aircraft into existing manned fleets provides immediately beneficial and realistic policies that may enhance wildfire control efforts \cite{ambrosia2011ikhana}. However, special attention must be paid to industry-standard aviation tactics, techniques, and procedures, and airspace models, to ensure collision avoidance \cite{albaker2009survey}.

Multi-agent problems with asynchronous actions and high-dimensional state spaces can be made tractable by using a hierarchical approach to planning \cite{ghavamzadeh2006hierarchical}. A valuable characteristic of heterogeneous multi-agent models is that action abstraction may be held constant by agent type. In our wildfire scenario, manned aircraft are provided optimal suppression locations but not guided to them. The process by which manned aircraft arrive at those suppression locations, through underlying primitive actions, is based on aerial firefighting standard operating procedures. Unmanned aircraft are explicitly guided to optimal surveillance locations through primitive actions that also satisfy collision avoidance requirements. A single manned aircraft action may occur over the course of multiple unmanned aircraft actions, resulting in an asynchronous action space. All aircraft share a single belief state, and possess each other's location. We decompose the initial attack multi-agent partially observable Markov decision process (MPOMDP) into separate surveillance and suppression partially observable Markov decision processes (POMDPs) operating on different time scales with distinct but related reward functions. The POMDPs are simplified into Markov decision processes (MDPs) using shared belief and uncertainty maps. The MDPs are subsequently coupled in a manner exploiting the known structure of an initial attack and respecting collision avoidance requirements. This framework may be adapted to optimize other aerospace teaming applications related to attack, search and rescue, and medical evacuation operations.

\indent
This paper's contributions include: (1) an adaptable, hierarchical framework for the manned and unmanned multi-agent, multi-objective, partially-observable initial attack problem featuring a high-dimensional state and action space, (2) several Monte Carlo tree search (MCTS) extensions informed by wildfire domain knowledge to include three suppression action space restriction mechanisms (ASRs), surveillance and suppression reward models, and an internal wildfire propagation model, (3) simulations demonstrating the efficacy of (1) and (2) given various wind conditions, elevation profiles, and resource topologies, and compared against firefighting techniques and a myopic baseline, and (4) a method for the early-dispatching of additional wildfire suppression assets to meet the 10 acre containment standard established by CALFIRE and other wildfire agencies. This framework, solver extensions, and methods validate the use of MDPs to optimize initial attack operations through the provision of collision avoidance, increased flexibility in suppression, and real-time wildfire insights.

\section{Related Work}
\noindent
This section reviews related research that applies distributed aircraft coordination frameworks, human-autonomous teaming models, hierarchical structures for multi-agent systems, or some combination of the three, to the wildfire surveillance and/or suppression problem. Further considered is the use of reinforcement learning or probabilistic search techniques to make informed decisions. We especially highlight the technical contributions of \citeauthor{seraj2021hierarchical}, which exist at the confluence of distributed aircraft coordination and hierarchical structures for multi-agent systems, for the joint wildfire surveillance and suppression problem \cite{seraj2021hierarchical}. Similarities and differences with our paper are discussed.

\noindent
\textbf{Distributed Aircraft Coordination Frameworks}

\noindent
The academic literature provides several examples of distributed unmanned aircraft coordination frameworks that support either wildfire surveillance or suppression (but rarely both). \citeauthor{julian2019distributed} demonstrate how deep reinforcement learning may be used to coordinate multiple autonomous fixed-wing aircraft to accurately track wildfire front expansion \cite{julian2019distributed}. \citeauthor{pham2018distributed} similarly introduce a distributed framework for controlling a set of quadcopters to chart a wildfire's progression while avoiding in-flight collisions \cite{pham2018distributed}. \citeauthor{ghamry2016cooperative} divide the wildfire surveillance problem into three stages: search, confirmation, and observation. In the search stage, the unmanned aircraft team uses a leader-follower approach and moves in geometric formation to the wildfire. On arrival, the unmanned aircraft distribute in accordance with a generated elliptical fire front perimeter \cite{ghamry2016cooperative}. \citeauthor{griffith2017automated} compare the use of MCTS and mathematical optimization (MO) as applied to the allocation of wildfire suppression teams. MO models the MDP as a mixed-integer linear optimization problem, then applies a commercial solver to determine feasible solutions \cite{griffith2017automated}.

\noindent
\textbf{Human-Autonomous Teaming Models}

\noindent
At its simplest, the relationship between man and machine in a system may be categorized as \textit{human-in-the-loop}, \textit{human-on-the-loop}, or \textit{human-out-of-the-loop}. Human-in-the-loop systems require human approval prior to action by the unmanned agent. Human-on-the-loop systems involve the human receiving updates from the unmanned agent, and the human in turn providing guidance to the unmanned agent. Human-out-of-the-loop systems have an unmanned agent that acts independently, albeit with initial or occasional guidance from the human \cite{hobbs2024safety}. There are surprisingly few research efforts that integrate man and machine in support wildfire management. The academic literature has largely prioritized fully autonomous teams conducting surveillance and suppression operations. There are exceptions. \citeauthor{bjurling2020drone} consider human-in-the-loop operator control over a swarm of unmanned wildfire surveillance aircraft \cite{bjurling2020drone}. Human-autonomous teaming may be defined as manned and unmanned agents working interdependently to accomplish a common goal \cite{o2022human}. Symbiotic autonomy is a form of teaming that accomplishes complex tasks by distributing sub-tasks and sharing information across multiple agents and agent groups \cite{basich2023competence}. In this framework, human and autonomous agents may act asynchronously to execute individual sub-tasks that enhance or inform each other's efforts \cite{veloso2015cobots}, and may be supported by a ``smart environment'' which provides shared understanding across the team and thereby improve performance. \citeauthor{seraj2020coordinated} introduce a distributed control framework for a team of unmanned aircraft conducting human-centered surveillance of wildfires and transmitting high-fidelity fire front observations to on-ground firefighters \cite{seraj2020coordinated}.

\noindent
\textbf{Hierarchical Structures for Multi-Agent Systems}

\noindent
A hierarchical representation of tasks can enable decision making across different levels of temporal abstraction for complex, multi-agent systems. Macro-actions are temporally extended actions that can be incorporated into multi-agent decision problems to overcome high-dimension state spaces. The use of macro-actions can result in cooperative agents acting asynchronously. \citeauthor{menda2018deep} introduce an algorithm that modifies policy gradient estimators for macro-actions, permitting policy optimization in models where agents act asynchronously. They successfully apply this algorithm to the cooperative multi-agent wildfire suppression problem \cite{menda2018deep}. Hierarchical representation is also a framework on which to scale reinforcement learning to large domains \cite{barto2003recent}. The category of hierarchical reinforcement learning (HRL) algorithms is large and diverse. With the advent of deep reinforcement learning, HRL has branched into  options \cite{bacon2017option} and sub-goal \cite{vezhnevets2017feudal} methods. Similarly, there exist several state-of-the-art multi-agent deep reinforcement learning (MARL) algorithms to include MADDPG \cite{lowe2017multi}, COMA \cite{foerster2018counterfactual}, and QMIX \cite{rashid2020monotonic}. An in-depth discussion of HRL and MARL algorithms is beyond of the scope of this paper. We do however highlight the work of \citeauthor{xu2023haven} in developing the HierArchical Value dEcompositioN (HAVEN) framework for solving decentralized partially-observable Markov decision process (Dec-POMDP) problems, one of the few algorithms straddling both HRL and MARL \cite{xu2023haven}.

\citeauthor{seraj2021hierarchical} apply a hierarchical framework of decision-making modules to a perception-action composite team of heterogeneous aircraft both surveying and suppressing a wildfire \cite{seraj2021hierarchical}. They model the environment as a multi-agent partially-observable Semi-Markov decision process (MAPOSMDP). A high-level module assigns specialized surveillance tasks to a set of perception UAVs, while a low-level module coordinates a control and planning framework through which all UAVs execute their assigned tasks. A novel reinforcement learning algorithm is proposed that employs a variant of the State-Action-Reward-State-Action (SARSA) algorithm tailored to multi-agent problems. This hierarchical design lends itself to ``prolific cooperation between perception and action agents''. We instead introduce a hierarchical framework that encourages cooperation between perception agents (unmanned surveillance aircraft), which in turn unilaterally support actions agents (manned suppression aircraft). Thus, the perception agents are a complement to the action agents. Action agents are modeled as non-cooperative interacting entities \cite{pierre2023multi}, and although provided suppression guidance, are otherwise uncontrolled. Ultimately, we seek to minimize the extent to which unmanned aircraft disrupt manned aircraft operations. We divide the initial attack MPOMDP formulation into surveillance and suppression decision processes operating on different time scales, and share information between their planners to ensure collision avoidance and fused information collection. Each sub-problem is resolved using MCTS with several domain-specific extensions. We emphasize the integration of manned aircraft conducting suppression operations, and demonstrate how certain tactics, techniques, and procedures employed by those aircraft can be supported, or at least avoided, by an unmanned aircraft network.

\section{Problem Statement}
\noindent
We begin by formulating the initial attack problem as a multi-agent partially observable Markov decision process (MPOMDP). The MPOMDP is a generalization of the MDP in which multiple agents, each unable to fully observe the underlying world state, collaborate and communicate freely towards one or more shared objectives \cite{kochenderfer2015decision}. The initial attack MPOMDP is represented by $\mathcal{M} = \langle \alpha, \mathcal{S, A}, P, O, \Omega, \gamma, R \rangle$, where $\alpha$ is the number of agents, $\mathcal{S}$ is the state space, $\mathcal{A}$ is the action space, $P$ is the transition model, $O$ is the observation function, $\Omega$ is the observation space, $\gamma$ is the discount factor, and $R$ is the reward function \cite{gmytrasiewicz2005framework}, \cite{pierre2023multi}. Tailoring $\mathcal{M}$ to the initial attack, take the number of agents to equal the number of participating unmanned and manned aircraft, or $\alpha$ = $\alpha_u$ + $\alpha_m$. Each possible state $s^t$ $\in$ $\mathcal{S}$ at time $t$ includes 1.) unmanned aircraft $i$ and manned aircraft $j$ positions and 2.) the wildfire state, such that $s^t$ = [p$_{i}^{t}$, p$^{t}_{j}$, W$^t$]. Action space $\mathcal{A}$ is the set of all possible surveillance actions for unmanned aircraft and all possible suppression actions for manned aircraft, giving $\mathcal{A}$ = ($\mathcal{A}_i$ $\times$ ... $\times$ $\mathcal{A}_{\alpha_u}$) $\times$ ($\mathcal{A}_j$ $\times$ ... $\times$ $\mathcal{A}_{\alpha_m}$). The transition model $P$ assigns the probability of transitioning from existing state $s^{t-1}$ by action $a$ to state $s^t$, with $\sum_{s \in \mathcal{S}} P(s^{t-1}, a, s^t) = 1$. The discount factor $\gamma$ is a hyperparameter that balances short and long-term rewards. A larger $\gamma$ prioritizes long-term rewards, whereas a smaller $\gamma$ prioritizes short-term rewards. Observation function $O$ gives a probability distribution over all possible observations $\omega$ after taking action $a$ resulting in state $s$. Observation space $\Omega$ includes the set of all possible partial wildfire observations by unmanned aircraft $i$ at time $t$, where $\omega_{i}^{t}$ = [W$_{i}^{t}$], $\omega^t$ = $\cup^{\alpha_u}_{i=1}$ $\omega_{i}^{t}$, and $\omega^t$ $\in$ $\Omega$. Each unmanned aircraft takes its own surveillance action for which it receives an individual observation. Observations are fused across all unmanned aircraft to attain shared observation $\omega^t$. The shared belief is a probability distribution over $\mathcal{S}$. We update the belief $b^t$ that the current state is $s^t$, for each $s^t$, using:

\begin{equation}
b^t(s^t) = \beta O(\omega^t, s^t, a^{t-1}) \sum_{s^{t-1} \in \mathcal{S}} b^{t-1}(s^{t-1}) P(s^t, a^t, s^{t-1})
\end{equation}
\noindent
where $b^{t-1}$ is the initial belief, $\omega^t$ is the new shared observation, and $\beta$ is a normalizing constant \cite{kaelbling1998planning}. Reward function $R$ returns the reward received when taking action $a$ from state $s$. Rewards are jointly considered across $\alpha$ aircraft.

The heterogeneity of the agent population suggests that the larger MPOMDP may be decomposed into smaller decision processes tailored to particular sub-tasks. We partition the initial attack agents into unmanned and manned agent groups, with each group possessing its own action space and a unique level of temporal abstraction. The initial attack itself delineates agent group responsibilities and defines agent group relationships. This permits the overarching MPOMDP to be divided into separable surveillance and suppression POMDPs with differing but linked actions spaces and reward models. The number of wildfire states is intractable at $2^{10,000}$, and a proxy in the form of a single modifiable belief map is introduced. The wildfire belief map size matches that of the actual wildfire state, and is updated in accordance with unmanned aircraft observations. Surveillance and suppression POMDPs are further simplified into MDPs by assuming the developed wildfire belief map is the actual wildfire state, which we find incurs an acceptable level of error. The resulting surveillance and suppression MDPs are building blocks in a hierarchical framework, and their associated planners repeat at a frequency specific to the wildfire environment. MDPs are P-Complete and finite horizon POMDP approximations are PSPACE complete, and the complexity of our approach is similar. This section introduces the stochastic wildfire propagation model, manned and unmanned aircraft dynamics, and state space formulations.

\subsection{Wildfire Propagation}
\begin{wrapfigure}{L}{0.5\textwidth}
\centering
\includegraphics[width=0.35\textwidth]{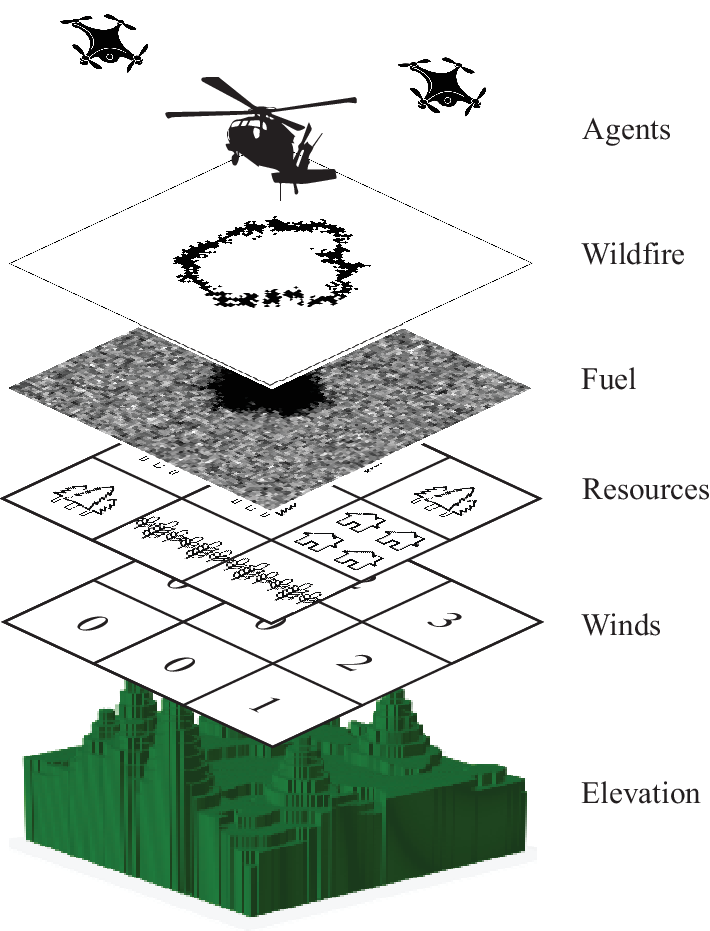}
\caption{The initial attack fire propagates based on fuel, winds, and terrain. Included in the initial attack environment are  aerial agents surveying and suppressing the wildfire, and the resources that are subject to destruction.}
\end{wrapfigure}
\noindent
The wildfire propagation problem has been the target of significant research activity, and the resulting models feature varying levels of complexity. We use a stochastic propagation model inspired by that of \citeauthor{bertsimas2017comparison} \cite{bertsimas2017comparison}, formulated to simulate the efficacy of the proposed hierarchical surveillance and suppression framework. The propagation model is modified to include the effects of aerial suppression, wind direction and strength, and terrain elevation \cite{papadopoulos2011comparative}. Wind and terrain are critical environmental factors affecting not only wildfire spread but unmanned and manned aircraft activity \cite{saikin2020wildfire}. In keeping with CALFIRE and other wildfire agencies, initial attack fire containment is  defined as wildfire control within two hours or spread not greater than $10$ acres. The wildfire model therefore consists of a roughly $10$ acre square discretized into a 100$\times$100 grid of 2$\times$2 meter cells. Each cell $x$ features a Boolean value representing whether $x$ is \textit{actually} on fire, $\mathcal{W}(x)$, a Boolean value representing whether $x$ is \textit{believed} to be on fire, $\mathcal{B}(x)$, and an integer value representing the amount of fuel remaining in $x$, $F(x)$. Fuel is defined as the amount of combustible material in a cell that contributes to fire behavior and effects \cite{landfirefuelbed}.

\indent
The duration of surveillance time-step $t$ is $d_t$, and is held constant at one minute. At $t$, each burning cell can either 1.) continue burning and expend a unit of fuel $\alpha$, 2.) stop burning by virtue of having expended all available fuel such that $F_t(x)$ = 0, or 3.) be partially or fully suppressed with probability $\delta(x)$. Similarly, each non-burning cell can ignite with probability $P$($x$), which is a function of the number of neighboring burning cells, remaining fuel, wind strength and direction, and terrain elevation. More specifically, the probability that neighboring cell $x'$ ignites cell $x$ at $t+1$ is $p(x, x')$.The duration of suppression time-step $T$ is $d_T$, equal to $kd_t$, where $k$ is a function of both the cruising speed of the suppression asset with and without load and the distance from the initial attack fire to a water replenishing source. There are five suppression actions, each with a different distributions of partially and fully suppressed wildfire grid cells. Notably, suppression affects not only the probability of ignition, but reduces fuel remaining as well \cite{hansen2012corrigendum}. The set of cells fully suppressed at $T$ with zero probability of ignition is $F_T$, where $\gamma_F$ is fuel reduced when fully suppressed. The set of cells partially suppressed at $T$ with probability of ignition compounded by $P_{partial}$ is $P_T$, where $\gamma_P$ is fuel reduced for a partially suppressed cell. The individual success probabilities of multiple suppressive actions on a single cell $x$ are independent. The resulting wildfire propagation equations are therefore

\begin{equation}
  F_{t+1}(x) =
    \begin{cases}
      \text{max(0, $F_t(x)$ $-$ $\alpha$ $-$ $\beta(x)$)} & \text{if $\mathcal{B}_t(x)$}\\
      \text{max(0, $F_t(x)$ $-$ $\beta(x)$)}            & \text{otherwise}
    \end{cases}
\end{equation}

\begin{equation}
  \beta(x) =
    \begin{cases}
      \text{$\gamma_F$} & \text{if $t \bmod k$ = 0 and x $\in$ $F_T$}\\
      \text{$\gamma_P$} & \text{if $t \bmod k$ = 0 and x $\in$ $P_T$}\\
      0         & \text{otherwise}
    \end{cases}
\end{equation}

\begin{equation}
  P(x) =
    \begin{cases}
      0                  & \text{$F_t(x)$ $=$ 0} \\
      \text{$\delta(x)$} & \text{if $\mathcal{B}_t(x)$ and $F_t(x)$ $>$ 0}\\
      \text{$\delta(x)$ (1 $-$ $\prod_{x'}$ (1 $-$ $p(x, x')$ $\mathcal{B}_t(x')$))} & \text{if $\neg\mathcal{B}_t(x)$ and $F_t(x)$ $>$ 0}
    \end{cases}
\end{equation}

\begin{equation}
  \delta(x) =
  \begin{cases}
    0                 & \text{if $t \bmod k$ = 0 and x $\in$ $F_T$} \\
    P_{partial} & \text{if $t \bmod k$ = 0 and x $\in$ $P_T$} \\
    1                 & \text{otherwise}
  \end{cases}
\end{equation}

\indent

Wind and elevation kernels may be convolved with a grid of $p(x, x')$ to help propagate the wildfire in the direction of the wind or towards up-sloping terrain. A modifiable resource grid is included in the initial attack environment to associate value to areas of importance, such as housing communities in the midst of a forest. The initial attack model begins with an assignment of fuel quantity and terrain elevation to each grid cell. As shown in Fig. 3, fire is seeded at $t$ = \SI{0}{\minute} at a handful of cells in the middle of the wildfire grid and allowed to propagate at each consecutive $t$. The simulation terminates when the initial attack becomes an escaped fire and additional suppression assets are introduced as a matter of policy; this occurs when $t$ = \SI{120}{\minute} or when the wildfire expands beyond the wildfire grid. The unmanned aircraft arrives five minutes after the initial attack fire begins, and the manned aircraft arrives 10 minutes after that. These estimates may be adjusted in the model depending on the proximity of assets to the wildfire.

\begin{figure*}[ht!]
\centering
\includegraphics[width=16.5cm]{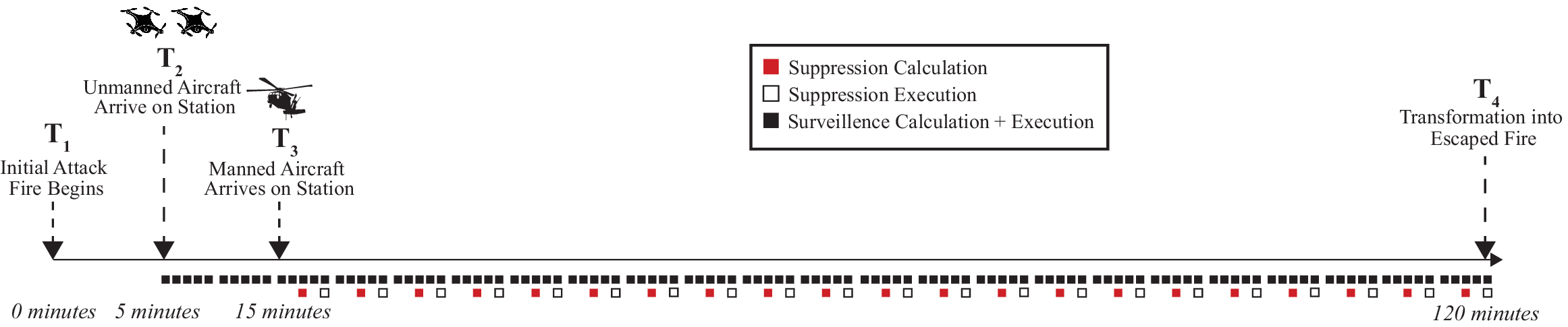}
\caption{The initial attack timeline depicts the arrival time for unmanned and manned aircraft, the escaped fire transformation, and the calculation and execution of surveillance and suppression actions.}
\end{figure*}

\subsection{Unmanned Aircraft Dynamics}
\noindent
Two multi-rotor unmanned aircraft with identical characteristics are introduced to survey the initial attack fire. The unmanned aircraft operate in a 10$\times$10$\times$7 airspace gridworld consisting of 20 $\times$20$\times$20 meter grid cells, and arrive on station at $t$ = \SI{5}{\minute}. The kinematics of each multi-rotor unmanned aircraft is abstracted in time, and simple gridworld commands, essentially primitive actions, translate the drone accordingly. More specifically, each drone may be translated up, down, left, right, ascend, descend, or hover in place. Translation may occur once per time-step $t$, and observations are collected en route to the next grid cell. Each drone is limited to the confines of the airspace gridworld, and the action space is pruned at the edges accordingly. Although each unmanned aircraft acts independently, the surveillance MPOMDP is simplified into a POMDP by having the controller take joint actions; by doing so, the surveillance action space increases from 7 to 49. A multi-rotor design was selected over a fixed-wing design to enable hovering, and because multi-rotor aircraft typically possess the kind of high-quality camera control needed to survey a wildfire.\\
\indent
Each unmanned aircraft must balance coverage with capture. The more elevated the unmanned aircraft, the wider its field of view, but the less resolution it has on the wildfire at a given location. Strategically, this results in a lower altitude selected when the fire is condensed, and a higher altitude when the fire is dispersed. The problem becomes more complicated with the addition of a second drone and collision avoidance penalties. As will be demonstrated, the inclusion of multiple unmanned aircraft that can actuate on two or more axes results in unique emergent behaviors, to include dispersion, loitering, stacking, and circling. Each unmanned aircraft is penalized by $P_u$ if within $D_u$ meters of another unmanned aircraft.


\subsection{Manned Aircraft Dynamics}
\begin{figure*}[ht!]
\centering
\includegraphics[width=16.5cm]{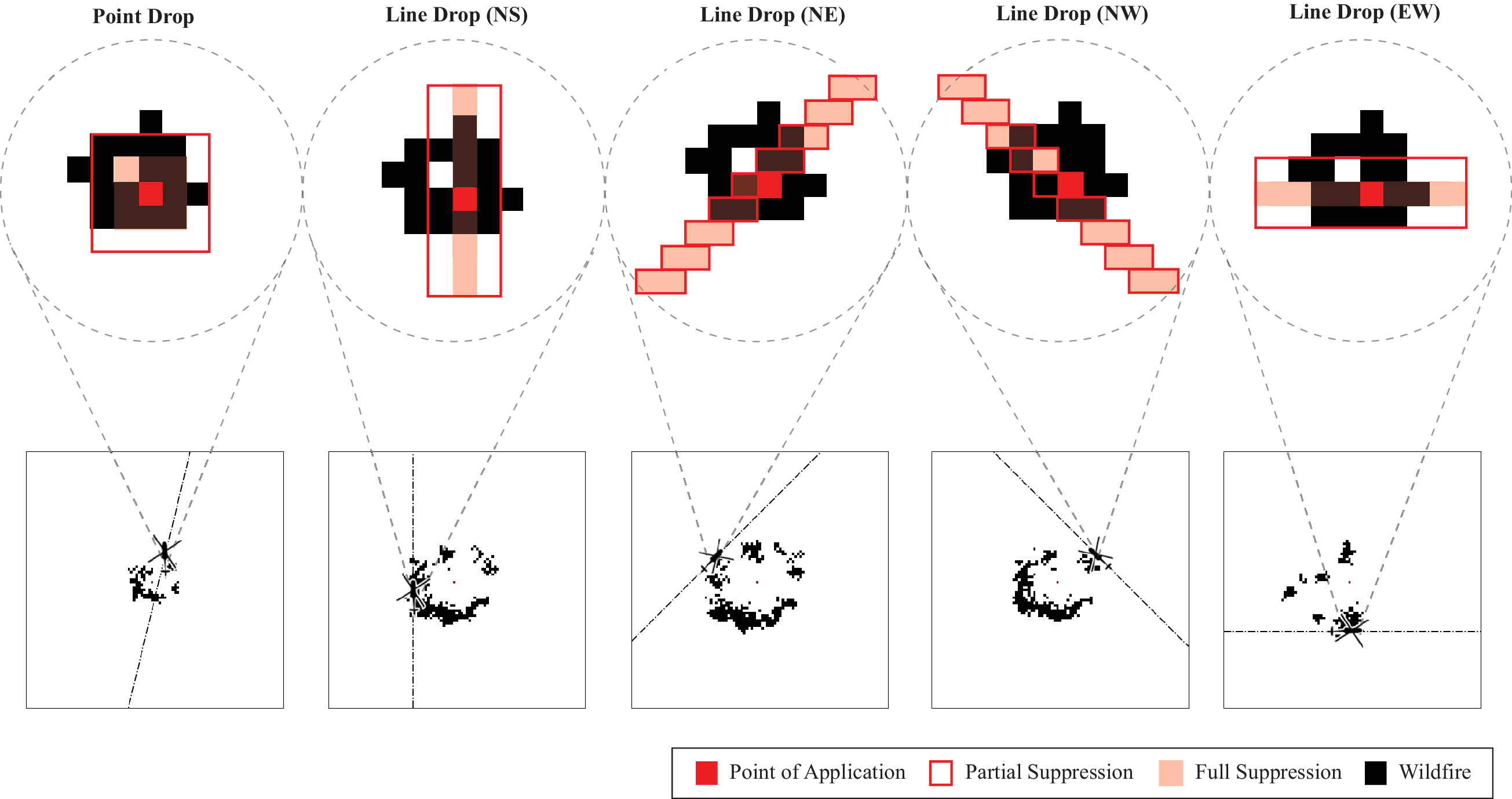}
\caption{Five suppression action drop-types and their associated aircraft axis of advance. A line drop and point drop have fundamentally different on-ground suppression characteristics.}
\end{figure*}
\noindent
A rotary-wing manned aircraft, with characteristics similar to a S-70 Firehawk helicopter, provides initial attack suppression capabilities through a short-line 660 gallon water bucket. The manned aircraft must fetch water from the nearest adequate water replenishing source prior to each suppression action. The distance between the water replenishing source and the initial attack fire can vary considerably, but is here assumed to be \SI{10}{\km}. Suppression time-step $T$ duration $d_T$ is tied to the distance between the water replenishing source and initial attack fire, and cruise speed of the manned aircraft with and without load. For example, a S-70 with an unloaded cruise speed of 140 KIAS and a loaded cruise speed of 80 KIAS, with \SI{10}{\km} between the initial attack fire and water replenishing source, can reasonably expect to perform a drop once every five minutes (thus, $d_T$ and $k$ = 5, where $k$ is the number of iterations that occur within $T$). Given an on station arrival at $t$ = \SI{15}{\minute}, the manned aircraft can therefore expect to perform 21 drops during the initial attack. These 21 water drops, in addition to efforts of ground-based suppression assets, determine whether an initial attack fire is contained or instead evolves into an escaped wildfire. Ground-based suppression activities are not considered in this paper.

The manned aircraft in an initial attack operates in accordance with best practices, in response to environmental factors, and often, per the guidance of a ground controller in the form of a fire truck, helitack crew, battalion chief, or hand crew \cite{sonoma2023}. Manned aircraft operations entail not only the placement of the water drop, but the type of drop, and aircraft axis of advance in light of winds, smoke, and other aircraft operating in the immediate vicinity \cite{kal2019efficiency}. This paper does not attempt to fully capture the decision making of a pilot in command of a firefighting aircraft, but rather, seeks to recognize the extent of that decision making, and provide supplemental guidance while ensuring supplemental assets (i.e. unmanned aircraft) do not restrict the manned aircraft's range of movement. Water may be dropped on any cell in the 100$\times$100 wildfire grid. As shown in Fig. 4, there are five standard drops, each categorized as either a point drop or line drop; this results in a suppression action space of 50,000. The point drop occurs when a suppressing helicopter slows to a near-hover prior to dumping the contents of its water bucket. The line drop occurs when a suppressing helicopter maintains forward airspeed prior to dumping the contents of its water bucket. Point drops are condensed in nature and typically used to suppress a condensed fire area, whereas line drops distribute the spread of water or retardant and are ideal for the placement of wet-lines. Drop selection, bucket line length, and bucket volume, determines the on-ground suppression profile \cite{ault2012drop}. This paper's model structure may be easily adapted to other aircraft and buckets of varying volumes and line lengths. \\
\indent
Simplifying assumptions include paralleling drop type and axis of advance, early identification of drop placement and type, and generalization of manned aircraft altitude and direction during the drop. The 200$\times$200 \si{\meter} wildfire grid is small relative to the operational area of a manned aircraft which may travel more than \SI{10}{\km} to reach a water replenishing source. It becomes reasonable to assume the manned aircraft's axis of advance is linked to its drop selection; a north-south line drop involves the manned aircraft traveling either north to south, or south to north, through the wildfire grid, such that the axis of advance includes the point of application. The axis of advance of a point drop is less straightforward, given the manned aircraft slows to a near-hover; however, it is assumed that the axis of advance parallels the vector extending from the water replenishing source to the point of application of the initial attack fire, such that additional maneuvering is not required and time is saved. Fig. 4 also depicts the axis of advance associated with each drop type. Each iteration of the surveillance model is held at one minute, providing the unmanned aircraft two minutes to determine suppression location and axis of advance. The temporal resolution may be increased to allow determination of manned aircraft maneuvering at a time nearer to suppression. Finally, for collision avoidance purposes, the unmanned systems are penalized by $P_m$ if they come within $D_m$ meters of the two-dimensional manned aircraft axis of advance at any altitude. This ensures that unmanned aircraft are penalized for proximity to the manned aircraft regardless of bucket-line length, drop altitude, or axis of advance directionality. Both $P_m$ and $D_m$ are larger than $P_u$ and $D_u$ respectively, representing the critical importance of ensuring the unmanned aircraft maintain separation from the manned aircraft, and the somewhat lesser importance of ensuring the unmanned aircraft maintain separation from one another.

Given the partial observability of the initial attack fire, a shared belief map $B$, matching the dimensions of the 100$\times$100 grid wildfire, is introduced and updated at regular intervals. The surveillance planner, executed at each time-step $t$, updates each wildfire cell in $B$ that is observed by either of the two unmanned aircraft, to reflect the actual wildfire state. The suppression planner, executed at each time-step $T$, may further update the belief and uncertainty maps at locations where suppression is assured. Fig. 5 depicts fuel and belief maps over the course of an initial attack fire. The red and green circular markers represent the unmanned aircraft locations in the two-dimensional representation, and the unmanned aircraft trajectory in the last five time-steps in the three-dimensional representation. The green square represents a high-value resource area. The ability of the unmanned aircraft to accurately capture the wildfire state is degraded as the wildfire expands.


\subsection{State Space Formulations}

\begin{figure*}[ht!]
\centering
\includegraphics[width=12cm]{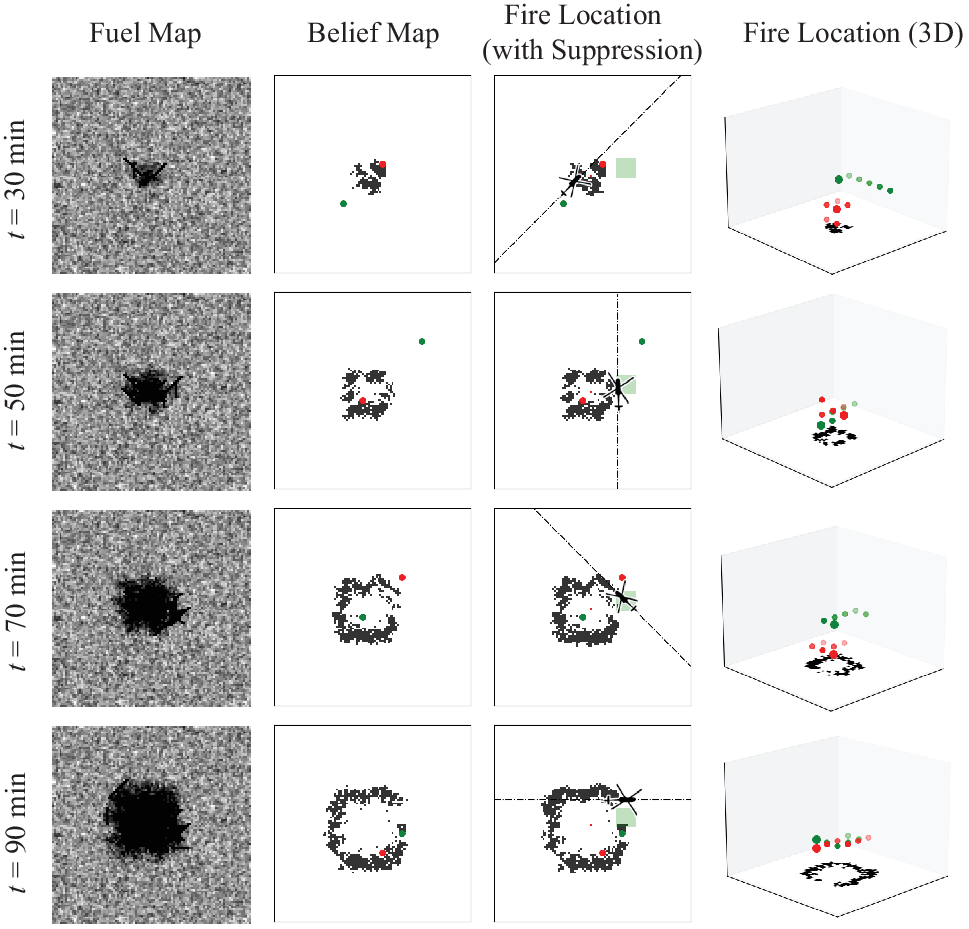}
\caption{Fuel, belief, and actual wildfire maps in two and three-dimensions as the initial attack fire propagates.}
\end{figure*}

\noindent
The surveillance POMDP is simplified into an MDP by assuming a shared belief map $\mathcal{B}$ to be the actual wildfire state rather than maintain a probability distribution across $2^{10,000}$ possible wildfire states; this does incur nontrivial error, but results in a notable reduction in computational complexity by detaching each instantiation of the evolving wildfire gridworld from the state space. Instead, the surveillance state space, which is explicitly defined, encompasses state variables ($X_{u_1}$, $Y_{u_1}$, $Z_{u_1}$, $X_{u_2}$, $Y_{u_2}$, $Z_{u_2}$). This results in a large, but manageable 490,000 states, equivalent to the Cartesian product of sets $U_1$ and $U_2$, each constituting all possible locations for an unmanned aircraft in the airspace. \\
\indent
The suppression problem also assumes the shared belief map $\mathcal{B}$ to be the actual wildfire state. Rather than an explicit formulation, the suppression state space is developed by calling a generative model on any number of state-action pairs ($s_0$, $a$). The state variables for the suppression MDP are ($X_{m}$, $Y_{m}$, $DR$, $2^{|P_T|}$). This generates a state space with an upper bound of approximately 13 billion states, representing every outcome of each possible suppression action.


\section{Solution}

\begin{figure*}[ht!]
\centering
\includegraphics[width=16.5cm]{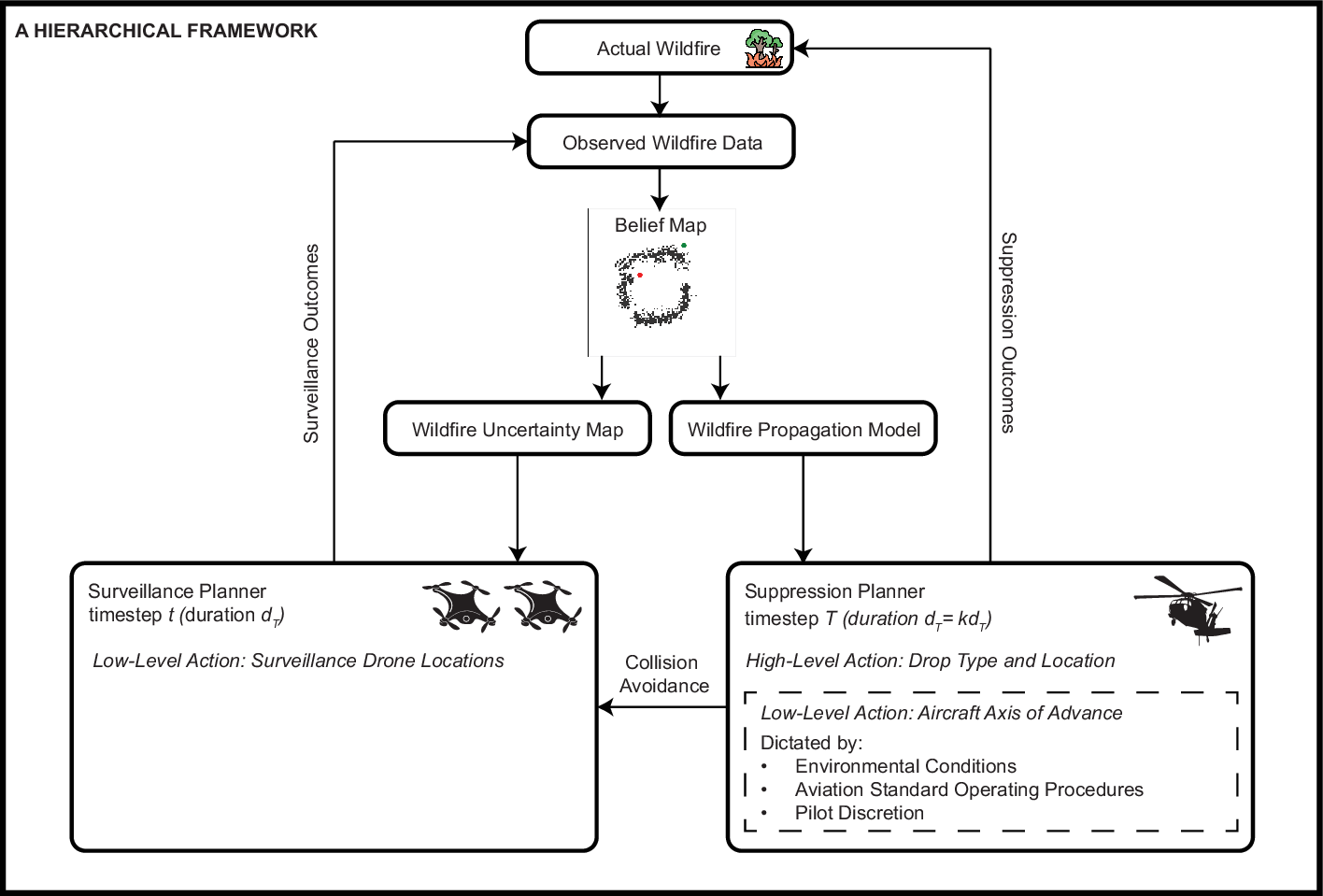}
\caption{A hierarchical framework for the initial attack using teaming involving linked surveillance and suppression planners.}
\end{figure*}
\noindent
A hierarchical planner derived from wildfire and teaming domain knowledge, shown in Fig. 6 and outlined in Algorithm 1, is used to split the larger surveillance-suppression MPOMDP into separate surveillance and suppression POMDPs operating on different time scales. A shared belief map is assumed for both models, further simplifying each POMDP into an MDP. The shared belief map is updated using the observed wildfire data, and used to generate a wildfire uncertainty map and in the internal wildfire propagation model. The wildfire uncertainty map informs the surveillance planner reward function. The surveillance planner recommends a surveillance action which updates the observed wildfire data and thus belief map. The internal wildfire propagation model informs the suppression planner rewards. The suppression planner recommends a suppression action, which affects the actual wildfire state. We choose to apply MCTS, a simulation-based search algorithm, to both surveillance and suppression planners. The Upper Confidence Bound for Trees (UCT) algorithm selects promising actions in the search trees \cite{kocsis2006bandit}. Several MCTS extensions are introduced to reduce the large surveillance and suppression action spaces. As shown in Algorithm 1, the execution of a planner-recommended action does not always immediately follow the planning process itself; this delay allows heterogeneous planners to adjust the recommendation of future actions based on the expected actions of another (e.g. in support of collision avoidance).

\begin{algorithm}
    \caption{Hierarchical Planner}
  \begin{algorithmic}[1]
    \INPUT Initial wildfire belief $\mathcal{B}_0$, surveillance state $\mathcal{S}_0$, uncertainty map $\mathcal{U}_0$
    \OUTPUT Updated $\mathcal{B}_k$, $\mathcal{S}_k$, $\mathcal{U}_k$ after $k$ iterations, suppression action $a$
    \FOR{$ i = 0, 1, ... , k-2 $}
      \STATE $e$ $\leftarrow$ Surveillance Planner ($\mathcal{U}_i, \mathcal{S}_i$)
      \STATE $\mathcal{B}_{i+1}$, $\mathcal{U}_{i+1}$, $\mathcal{S}_{i+1}$ $\leftarrow$ survey($\mathcal{B}_i, \mathcal{U}_i, e$)
    \ENDFOR
    \STATE $a$ $\leftarrow$ Suppression Planner ($\mathcal{B}_{k-1}$)
    \STATE $e$ $\leftarrow$ Surveillance Planner ($\mathcal{U}_{k-1}$, $\mathcal{S}_{k-1}$, $a$)
    \STATE $\mathcal{B}_{k}$, $\mathcal{U}_{k}$, $\mathcal{S}_{k}$ $\leftarrow$ survey($\mathcal{B}_{k-1}, \mathcal{U}_{k-1}, e$)
    \STATE $\mathcal{B}_{k}$ $\leftarrow$ suppress($\mathcal{B}_{k}, a$)
  \end{algorithmic}
\end{algorithm}


\subsection{MCTS Solver}
\noindent
The introduced hierarchical planner is solver-agnostic. That said, the large state and action spaces coupled with a non-stationary environment in the propagating wildfire encourages use of an online, stochastic planning algorithm like MCTS. MCTS is also an anytime algorithm, which means it can return a valid solution if interrupted before runtime completion \cite{browne2012survey}. The set of possible actions are represented as edges, and resulting states as nodes, in the search tree. Each MCTS iteration follows four general steps: selection, expansion, simulation, and propagation. MCTS selects nodes and traverses the search tree using a predefined heuristic informed by the problem domain. This heuristic results in an asymmetric search where the most promising actions are prioritized. We choose to apply the standard Upper Confidence bound for Trees (UCT) algorithm to balance exploration versus exploitation in the tree search policy and determine the value of each node $n$, as follows
\begin{equation}
  \text{UCB}(n) = \bar{u}(n) + c \sqrt{\frac{\text{log(visits}(n))}{\text{visits}(n')}}
\end{equation}
such that $\bar{u}(n)$ is the value of the state at $n$, $c$ is an exploration constant that adjusts the balance between exploration of new nodes and exploitation of previously visited nodes, visits($n$) is the visitation count for $n$, and visits($n'$) is the visitation count for the parent node of $n$, $n'$. The search tree is then expanded by adding a child node to the selected node. State value estimation occurs by simulating a play-out from a node to the end of a predefined planning horizon. A random policy is here applied to estimate action value during search tree roll-outs. The simulation results are then backpropagated up the search tree, concluding a search iteration. It can be difficult to accurately assess MCTS computational complexity given the many sub-tasks involved, but in the general case, MCTS runtime and memory complexity scale linearly with the number of search iterations \cite{li2023self}, \cite{powley2017memory}. Fig. 7 shows how MCTS may be applied to search trees for unmanned aircraft surveying an initial attack wildfire. Selection is depicted by the bold circles, and involves traversing the tree to a select depth using an algorithm like Upper Confidence bounds applied to Trees. The selected node is then expanded by adding a child node, shown in grey. A simulation is preformed from the child node to a preordained depth or condition, shown in red. The simulation results are associated with the child node, and backpropagated up the search tree. This repeats until a given condition is true, computation time-limit is met, or a ceiling on search iterations is reached. Given that the state space is evolving in time, we use an internal wildfire model to propagate the wildfire belief map between progressive search tree depths. States that exceed the boundaries of the model, shown crossed out in red, are pruned from the search tree.

\begin{figure*}[ht!]
\centering
\includegraphics[width=16.5cm]{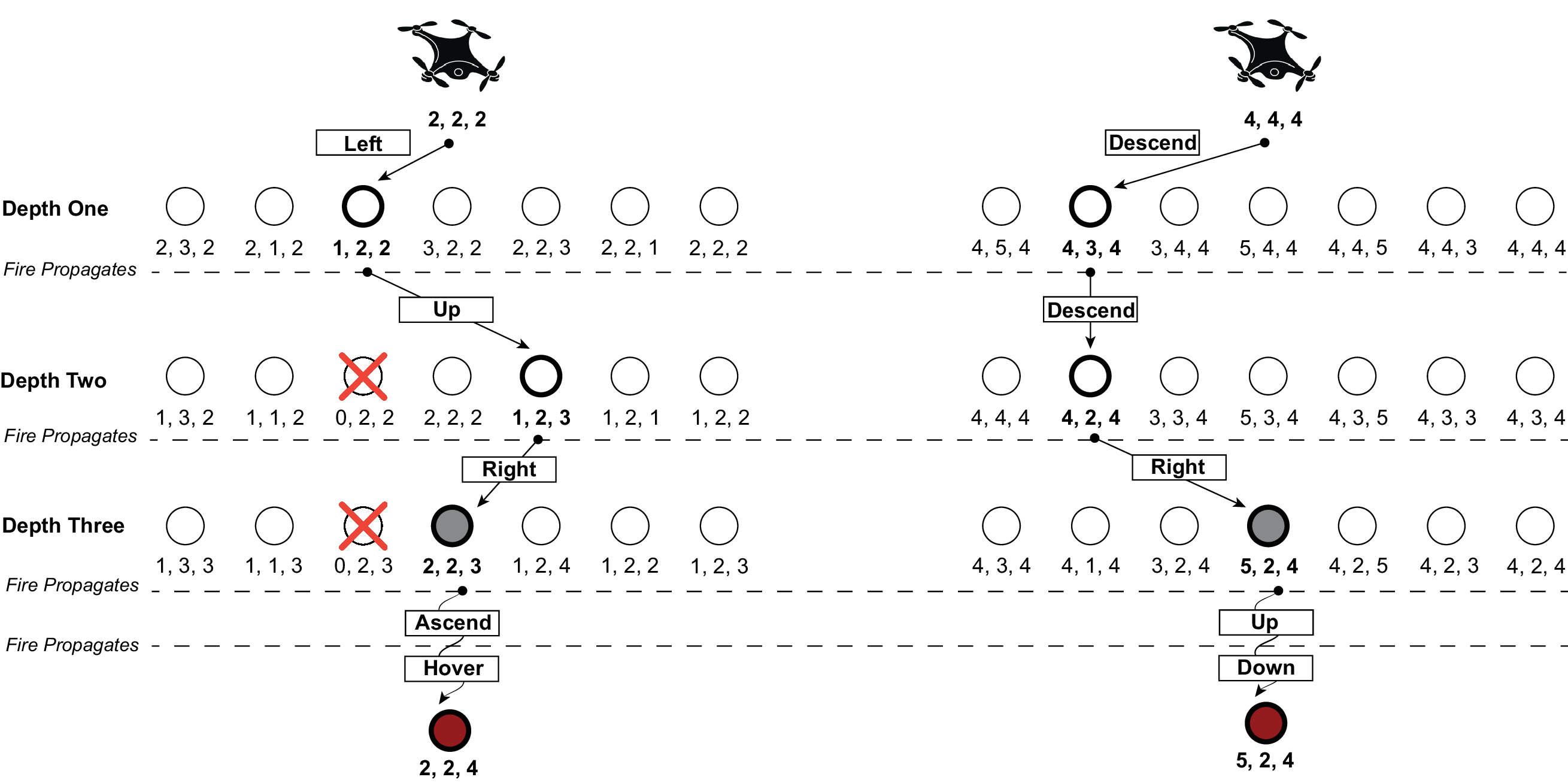}
\caption{Monte Carlo search trees depicted for two unmanned aircraft surveying an initial attack wildfire. Each iteration of Monte Carlo tree search has four steps: selection, expansion, simulation, and back-propagation.}
\end{figure*}

\subsection{Surveillance Planning}

\begin{algorithm}
    \caption{[Simplified] Surveillance Planner (Uncertainty Reward Model)}
  \begin{algorithmic}[1]
    \INPUT Uncertainty map $\mathcal{U}_t$, surveillance state $\mathcal{S}_t$, suppression action $a$ (optional)
    \OUTPUT Surveillance action $e$
    \STATE \textbf{Initialize} Surveillance action score map $\tilde{\mathcal{E}}$ $\leftarrow$ $\emptyset$
    \FOR{ action $e_i$ $\in$ $E$}
      \STATE $S_{t+1}$ $\leftarrow$ apply $e_i$ to $S_t$
      \STATE $\mathcal{V}_{e_i}$ $\leftarrow$ ranging($S_{t+1}$)
      \STATE $R_{o_{e_i}}$ $\gets$ $\Sigma_{x \in \mathcal{V}_{e_i}} \mathcal{U}_t(x)$
      \STATE $R_{U_{e_i}}$ $\gets$ $\tau_1$ $R_{o_{e_i}}$ + $\tau_2$ $P_u(S_{t+1})$ + $\tau_3$ $P_m(S_{t+1}, a)$ + $\tau_4$ $P_i(S_{t+1})$
      \STATE $\tilde{\mathcal{E}}[e_i]$ $\leftarrow$ $R_{U_{e_i}}$
    \ENDFOR
    \STATE Recommended surveillance action $e$ $\leftarrow$ argmax$_{e_i}$ $\tilde{\mathcal{E}}[e_i]$
  \end{algorithmic}
\end{algorithm}

\noindent
There exists 49 possible surveillance actions at time $t$ for a given belief map, equivalent to the Cartesian product of the set of seven actions for two unmanned aircraft. At depth of two, there becomes 2401 possible actions, and at depth three, there are 1.18$\times$$10^5$. Given 2$^{10,000}$ wildfire states, developing an explicit policy using dynamic programming is infeasible. Instead of an explicit policy formulation, a sampling-based approach with a generative model is used. An additional complication is reward uncertainty, given a stochastic element in the selection-process for observations following a selected surveillance action. As each unmanned aircraft increases in altitude, the resolution of the wildfire grid below decreases, and a sampling process is used to balance observation coverage and capture.

Algorithm 2 outlines a process by which a surveillance action with depth one is recommended; as will be later discussed, a probabilistic-search formulation of this process is applied to enable search depths of two and three. Uncertainty map $\mathcal{U}_t$, surveillance state $\mathcal{S}_t$, and optionally, suppression action $a$ are submitted to the surveillance planner. A surveillance score map $\tilde{\mathcal{E}}$ is initialized. Each surveillance action $e$ in set $E$ is applied to the current surveillance state $\mathcal{S}_t$ to attain the next surveillance state $\mathcal{S}_{t+1}$. The ranging function provides an array of wildfire grid cells $\mathcal{V}_{e}$ which are observed from the updated surveillance state position, as a function of both location and altitude. As mentioned, there is a stochastic element to the ranging function, meaning that observations attained from a particular surveillance position, and thus ensuing rewards, will differ slightly with each simulation. The uncertainty in observation for each wildfire cell $x$ in set $\mathcal{V}_{e}$, $U(x)$, is then summed to attain $R_{o_{e}}$, one of four components of the total surveillance reward $R_{U}$ for a particular action $e$. This uncertainty model thus provides a reward corresponding to the extent to which observations reduce overall belief map uncertainty, regardless of the actual state of the wildfire or belief map. As shown in Fig. 8, the uncertainty reward model maintains a 100$\times$100 uncertainty map in which each cell not observed in a particular time-step increments its uncertainty by an amount proportional to its proximity to burning cells (in the belief map), and each cell observed resets its uncertainty to zero. Cells on the outskirts of a wildfire grid with negligible likelihood of igniting change their uncertainty minimally, while cells at the head of the fire change their uncertainty rapidly. Observations of cells that have not been recently observed reward more than cells that were more recently observed, regardless of whether the cell changes state post-observation.

\begin{figure}[ht!]
 \centering
 \includegraphics[width=16.5cm]{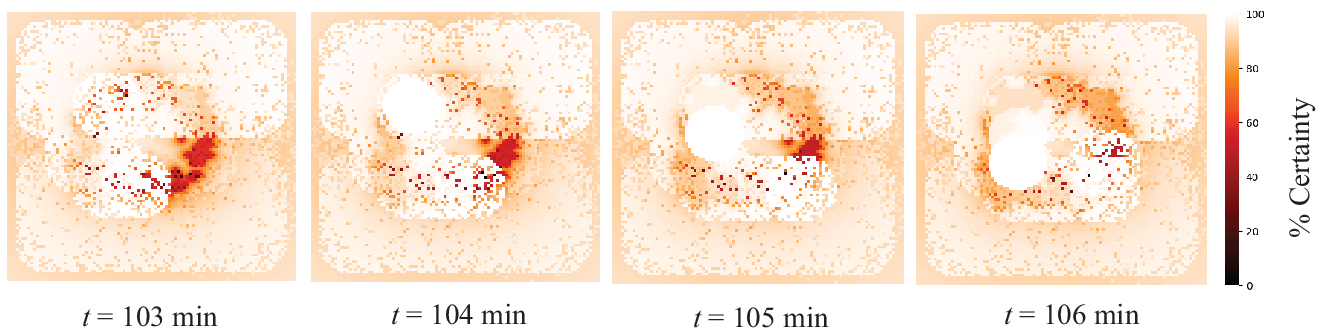}
 \caption{Wildfire uncertainty map during initial attack fire propagation. Dark red areas suggest nearby fire activity but few recent observations. Grid cells distant from the wildfire, such as at the map periphery, increment their percent uncertainty minimally.}
\end{figure}

Reward $R_U$ is further comprised of penalties $P_u$, $P_m$, and $P_i$. Unmanned aircraft $U_1$ and $U_2$ are penalized for their proximity to one another ($P_u$) and to the manned aircraft's axis of advance ($P_m$). The axis of advance and aircraft direction along it may be inferred from environmental data such as winds, aviation and wildfire best practices, and the suppression action $a$, or alternatively, they may be explicitly passed to the surveillance planner. A small penalty ($P_i$), proportional to the distance from each unmanned aircraft to the initial attack fire origin $IA_O$, is added to encourage unmanned aircraft to approach the initial attack fire. Penalty $P_i$ is quickly overtaken by other rewards and penalties. The resulting total reward equation for the unmanned aircraft follow, where $\tau_1$, $\tau_2$, $\tau_3$, and $\tau_4$ are tunable parameters:

\begin{equation}
  R_o = \tau_1 \sum_{x \in \mathcal{V}_{e}} \mathcal{U}_t(x)
\end{equation}
\begin{equation}
  P_u =
    \begin{cases}
      \tau_2 & \text{if $\|U_1$ $-$ $U_2\|$ $\le$ $D_u$}\\
      0 & \text{otherwise}\\
    \end{cases}
\end{equation}
\begin{equation}
  P_m =
    \begin{cases}
      \tau_3 & \text{if $\|U_1$ $-$ $AOA\|$ $\le$ $D_m$}\\
      \tau_3 & \text{else if $\|U_2$ $-$ $AOA\|$ $\le$ $D_m$}\\
      0 & \text{otherwise}\\
    \end{cases}
\end{equation}
\begin{equation}
  P_i = \tau_4 (\lVert U_1 - IA_O \rVert + \lVert U_2 - IA_O \rVert)
\end{equation}
\begin{equation}
  R_U = R_o - P_u - P_m - P_i
\end{equation}

The simplified approach in algorithm 2 becomes computationally intractable when considering depths greater than one given the large action space and reward uncertainty. The algorithm 2 reward structure is maintained, and MCTS with UCT is applied to a generative MDP model to search the action space efficiently. The wildfire belief state $\mathcal{B}_t$ is passed to the MDP model to be adjusted and modified, along with uncertainty map $\mathcal{U}_t$, with progressive extension of the search-tree's depth. Each MCTS surveillance calculation is capped at \SI{30}{\second} of calculation time, to allow each drone to proceed to its new location and attain observations prior to the next iteration. Any action which would take either drone beyond gridworld boundaries, or any combination of actions which would take both drones into the same grid cell, is pruned. The computational complexity of a single MCTS surveillance iteration increases with wildfire growth. For example, given a \SI{30}{\second} restriction, and depth of three, MCTS surveillance conducts roughly 800 iterations at $t$ = \SI{0}{\minute}, and only 60 iterations at $t$ = \SI{120}{\minute}. Reducing the MCTS depth as the wildfire progresses ensures a roughly equivalent number of iterations at depth of one, throughout the lifespan of the wildfire. A depth factor of three is reduced to two, and then one, as the wildfire propagates, to maintain a minimum of two runs for each possible action at depth of one.

\subsection{Suppression Planning}

\begin{algorithm}
    \caption{[Simplified] Suppression Planner (Localized Destruction Minimization Reward Model)}
  \begin{algorithmic}[1]
    \INPUT Wildfire belief $\mathcal{B}_t$, hyperparameter set $\{Q, RO\}$
    \OUTPUT Suppression action $a$
    \STATE $\mathcal{B}_{t+RO}$ $\leftarrow$ propagate($\mathcal{B}_{t}$, $RO$)
    \STATE $A$ $\leftarrow$ ASR($\mathcal{B}_{t}$, $Q$)
    \STATE \textbf{Initialize} Suppression action score map $\tilde{\mathcal{A}}$ $\leftarrow$ $\emptyset$
    \FOR{action $a_i \in A$}
      \STATE where $a_i$ is $(X_m, Y_m, DR)$
      \STATE $\mathcal{B}_{st}$ $\leftarrow$ suppress($X_m$, $Y_m$, $DR$, $\mathcal{B}_{t}$)
      \STATE $\mathcal{B}_{st+RO}$ $\leftarrow$ propagate($\mathcal{B}_{st}$, $RO$)
      \STATE $\mathcal{L}_s$ $\leftarrow$ localize($\mathcal{B}_{st+RO}$, $X_m$, $Y_m$)
      \STATE $\mathcal{L}$ $\leftarrow$ localize($\mathcal{B}_{t+RO}$, $X_m$, $Y_m$)
      \STATE $R_{M_i}$ = $\Sigma_{x \in \mathcal{L}} \mathcal{D}(x) - \Sigma_{x \in \mathcal{L}_s} \mathcal{D}(x)$
      \STATE $\tilde{\mathcal{A}}[a_i]$ $\leftarrow$ $R_{M_i}$
    \ENDFOR
    \STATE Recommended suppression action $a$ $\leftarrow$ argmax$_{a_i}$ $\tilde{\mathcal{A}}[a_i]$
  \end{algorithmic}
\end{algorithm}

\noindent
The suppression model features 50,000 possible actions, encompassing five suppression actions centered on each of 10,000 wildfire grid cells, for a given belief map. The large action space again prevents an exact policy formulation, and a generative method is instead applied. Reward uncertainty remains a factor, as water buckets fully suppress some wildfire cells, but only partially suppress others; the partially suppressed cells are selected via random sampling. An understanding of wildfire dynamics allows for various forms of ASR that minimize the overall action space by 99$\%$ or more with minimal consequence. A generated and internally-held wildfire propagation mode with limited information, shown in Fig. 6, enables the suppression planner to conduct customized roll-outs to optimize action selection. Two reward models are here introduced, localized and global resource destruction minimization. Global resource destruction minimization calculates the instantaneous destruction of the full 100$\times$100 grid after roll-out, whereas localized destruction minimization calculates the instantaneous destruction of a smaller grid centered on the location of the action taken after roll-out. In both cases, the suppression aircraft is rewarded $R_M$, or penalized $P_M$, proportional to the instantaneous destruction caused by the wildfire following one or more roll-out periods in which the internal fire propagation model with limited information propagates post-suppression. Instantaneous destruction is a function of number of cells burning and the value of resources contained within those cells. Rewards for minimized instantaneous resource destruction following a roll-out period ensures the long-term impact of water lines can be measured against the short-term impact of more immediate suppressive activities.

Localized resource destruction minimization examines only the area immediately surrounding the suppression activity for reward considerations. Suppressive activities applied to different portions of the wildfire cannot be directly compared due to the differing windows under consideration, and must instead be compared against a non-suppressed but equally propagated ``reference grid'', as illustrated in Fig. 9. Reward is then maximized when the difference between the appropriately localized portions of the reference grid and suppressed grid are greater. Algorithm 3 outlines the process by which a suppression action $a$ is selected using localized rewards. Wildfire belief map $\mathcal{B}_t$ is propagated to roll-out depth $RO$ to attain propagated reference grid $\mathcal{B}_{t+RO}$. An internal wildfire propagation model possessing wind and elevation, but only limited fuel, data is used. While the quality of the internal propagation model affects reward optimization, quality results may be attained for even a low-fidelity model. Belief map $\mathcal{B}_t$ is also submitted to one of three ASR functions with quantile selection $Q$ to attain a smaller action space set $A$. Each suppression action $a_i$ in set $A$ is then applied to $\mathcal{B}_t$, the results of which are propagated to depth $RO$  to attain $\mathcal{B}_{st+RO}$. A grid with size proportional to $RO$ is centered on action location ($X_m$, $Y_m$), and corresponding subarrays $\mathcal{L}$ and $\mathcal{L}_s$ are indexed from $\mathcal{B}_{t+RO}$ and $\mathcal{B}_{st+RO}$ respectively. The summed instantaneous destruction in both subarrays are calculated and subtracted from one another to attain reward $R_{M_i}$.

\begin{figure}[ht!]
 \centering
 \includegraphics[width=16.5cm]{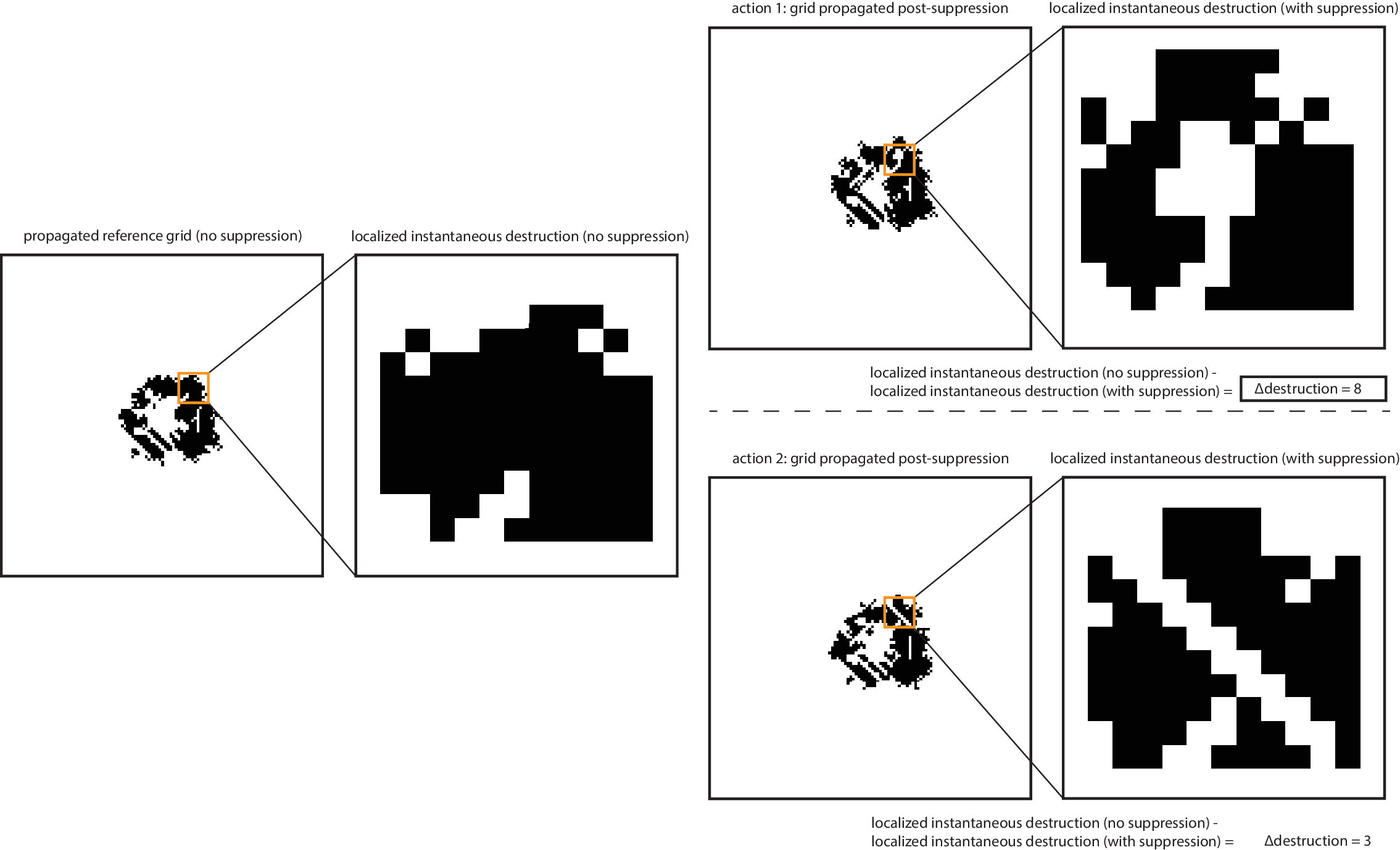}
 \caption{The localized suppression destruction minimization reward model. A propagated reference grid and post-suppression propagation grid are clipped to attain local subarrays which are then compared to determine action consequence.}
\end{figure}

\begin{algorithm}
    \caption{[Simplified] Suppression Planner (Global Destruction Minimization Reward Model)}
  \begin{algorithmic}[1]
    \INPUT Wildfire belief $\mathcal{B}_t$, hyperparameter set $\{Q, RO\}$
    \OUTPUT Suppression action $a$
    \STATE $A$ $\leftarrow$ ASR($\mathcal{B}_{t}$, $Q$)
    \STATE \textbf{Initialize} Suppression action score map $\tilde{\mathcal{A}}$ $\leftarrow$ $\emptyset$
    \FOR{action $a_i \in A$}
      \STATE where $a_i$ is $\{X_m, Y_m, DR\}$
      \STATE $\mathcal{B}_{st}$ $\leftarrow$ suppress($X_m$, $Y_m$, $DR$, $\mathcal{B}_{t}$)
      \STATE $\mathcal{B}_{st+RO}$ $\leftarrow$ propagate($\mathcal{B}_{st}$, $RO$)
      \STATE $P_{M_i}$ $\gets$ $\Sigma_{x \in \mathcal{B}_{st+RO}} \mathcal{D}(x)$
      \STATE $\tilde{\mathcal{A}}[a_i]$ $\leftarrow$ $P_{M_i}$
    \ENDFOR
    \STATE Recommended suppression action $a$ $\leftarrow$ argmin$_{a_i}$ $\tilde{\mathcal{A}}[a_i]$
  \end{algorithmic}
\end{algorithm}

Global resource destruction minimization, shown in Fig. 10, allows for the direct comparison of any action taken across the global grid following the appropriate propagation sequence. However, the scope of destruction often causes the impact of the selected suppression activity to get lost amid the extent of the stochastic fire spread. Algorithm 4 outlines the process by which a suppression action $a$ is selected using global rewards. Unlike the localized approach, the global reward model does not initialize a single non-suppressed, propagated reference grid at the onset. Instead, action roll-out rewards are compared directly. The belief map $\mathcal{B}_t$ is again submitted to one of three ASR functions with quantile selection $Q$ to attain a smaller action space set $A$. Each suppression action $a_i$ in set $A$ is then applied to $\mathcal{B}_t$, the results of which are propagated to depth $RO$ to attain $\mathcal{B}_{st+RO}$. The summed instantaneous destruction of $\mathcal{B}_{st+RO}$, now global rather than localized, is used to attain penalty $P_{M_i}$.

\begin{figure}[ht!]
 \centering
 \includegraphics[width=10cm]{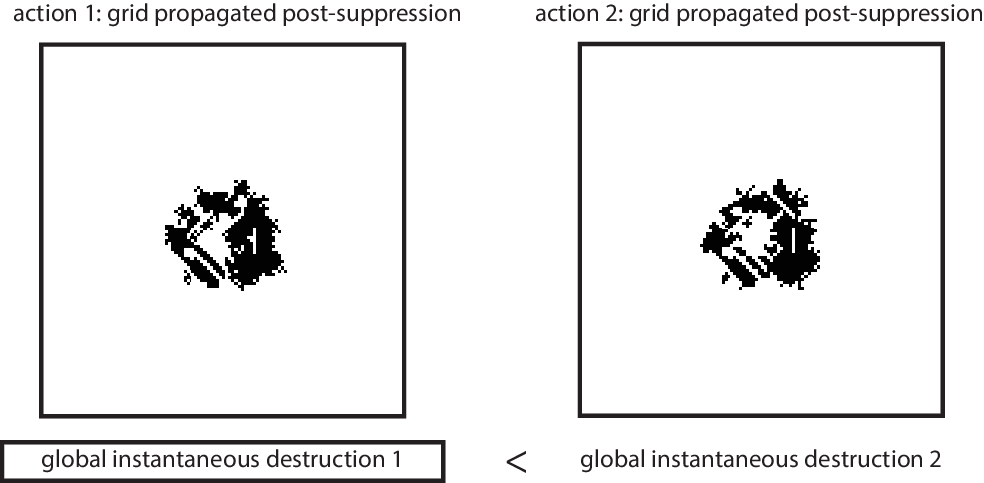}
 \caption{The global suppression destruction minimization reward model. Unlike in the localized model, there is no propagated reference grid. The wildfire belief map is suppressed, then propagated, then compared across suppression actions.}
\end{figure}

Three ASR methods are presented. ASR method 1 limits suppression actions to cells burning in the belief map; it can be assumed that ideal suppression locations involve at least one burning cell. ASR method 1 reduces the overall action space by 90-95\%. ASR method 2 restricts suppression to a pre-determined percentage of cells, here held at 5, 10, or 15\%, that are furthest away from the fire origin. ASR method 2 recognizes that suppressing the wildfire exterior, and wildfire head, results in the greatest likelihood of minimizing overall wildfire propagation. ASR method 2 reduces the overall action space by 99\% or more. ASR method 2 may be modified to increase the percentage of considered wildfire cells, the set of which is then randomly selected from. A stochastic wildfire does not spread evenly, and consideration of cells to include those not furthest away may better suppress multiple fire regions. ASR method 3 further restricts ASR method 2 to two 60 degree arcs, one centered on the area of highest resource value, and the other on the wildfire head. Cells qualified by ASR method two that are also within the established arcs, qualify for suppression per ASR method 3. These arcs may shift as environmental factors change throughout the simulation.

\begin{figure*}[ht!]
\centering
\includegraphics[width=16.5cm]{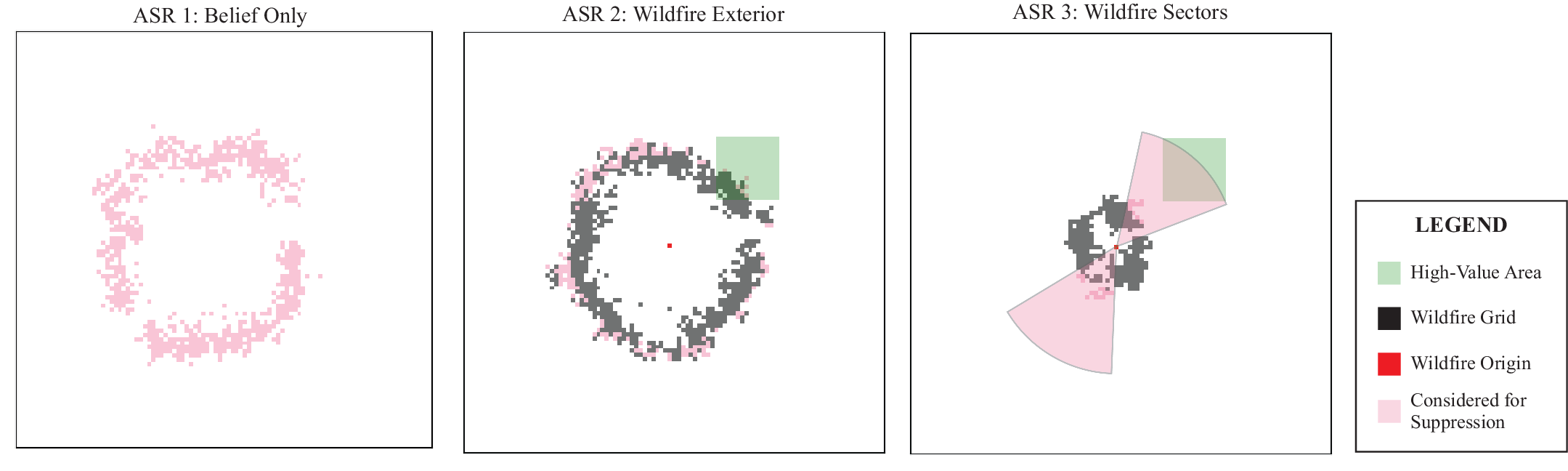}
\caption{Three action space restriction schemes: belief only, wildfire exterior, and wildfire sectors based on high-value areas and the wildfire head.}
\end{figure*}

MCTS with UCT is again applied to a generative MDP model with the algorithms 3 or 4 used to determine action rewards. Each MCTS suppression calculation is capped at \SI{120}{\second} of calculation time, to allow the suppression calculation updated belief information while also permitting time to execute the suppression action prior to the next iteration.

\subsection{Early Dispatch}
\noindent
The development of an intertwined surveillance-suppression model enables the early allocation of additional suppression assets in instances where wildfire growth exceeds even the optimized capabilities of a single  aircraft. Challenges include when to consider requesting additional assets, and what thresholds should determine that additional assets are required. The vast majority of initial attack fires do not become escaped fires, and premature dispatching of aircraft is therefore costly and wasteful. At the same time, initial attack fires destined to become escaped fires have limited windows of opportunity in which additional suppression can result in containment; after a certain point, the wildfire spread is too significant to be reasonably contained except by a full-blown suppression operation. A regression-backed approach is introduced in Algorithm 5, wherein the wildfire ring radius at $t$ = \SI{120}{\minute} is predicted at each time-step $T$ following suppression execution. If the predicted wildfire ring radius exceeds a given time and ring threshold, then a second S-70 is dispatched to the scene to conduct suppression activities.

\begin{algorithm}
    \caption{Early Dispatch Procedures}
  \begin{algorithmic}[1]
    \INPUT Wildfire belief $\mathcal{B}_t$, historical wildfire ring array $\mathcal{R}$
    \OUTPUT Early dispatch recommendation
    \IF {$t$ mod $d_T$ = 0}
    \STATE $\mathcal{R}_t$ $\leftarrow$ radius($\mathcal{B}_t$)
    \STATE $\mathcal{R}$[$t$] $\leftarrow$ $\mathcal{R}_t$
    \STATE $\mathcal{R}_{t=120}$ $\leftarrow$ regression($\mathcal{R}$)
    \IF {($t$ > time threshold) and ($\mathcal{R}_{t=120}$ > ring threshold)}
        \STATE Recommend early dispatch
    \ELSE
        \STATE Recommend against early dispatch
    \ENDIF
    \ENDIF
  \end{algorithmic}
\end{algorithm}

\section{Experimentation}
\subsection{Environmental Models}
\noindent
\textbf{Abstracted Case Studies}
\\
\noindent
Three abstracted case studies are presented to simulate the introduced hierarchical framework, surveillance and suppression reward models, and MCTS extensions in various environmental conditions. Case 1 involves flat terrain, variable winds, and a single high-value resource area. Case 2 again features flat terrain, has two high-value resource areas, but also experiences a significant but randomized wind-shift at $t$ = \SI{60}{\minute}. Case 3 has hilly to mountainous elevation, variable winds, and three high-value resources areas. Each progressive case features increased opportunity for destroyed resources.

\begin{figure}[ht!]
\centering
\includegraphics[width=16.5cm]{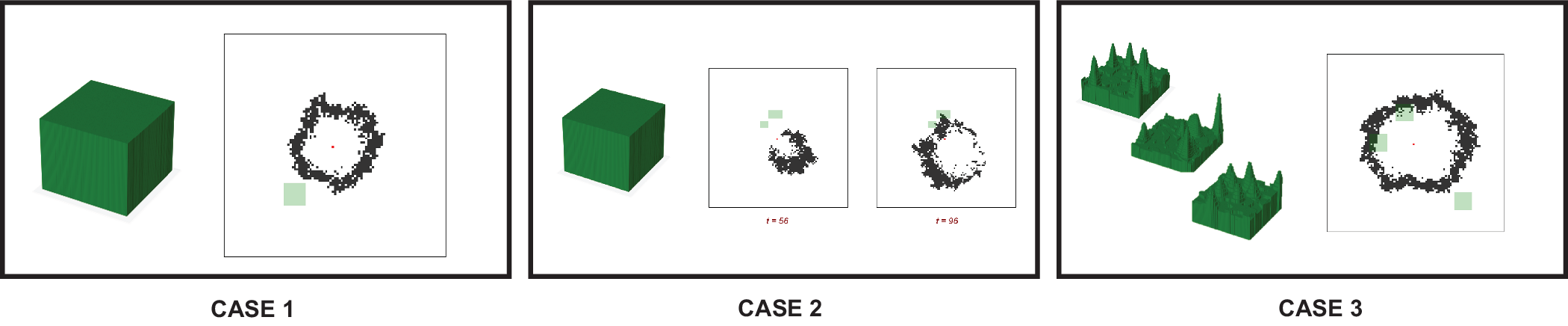}
\caption{Three abstracted environmental models featuring different terrain elevation maps, environmental wind profiles, and number and size of high-value areas.}
\end{figure}

\noindent
\textbf{Actual Wildfire Case Study} \\
The state of Hawaii is particularly susceptible to wildfires due to an abundance of fire-prone grasses and shrubs and an increasingly warm and dry climate \cite{ellsworth2014invasive}, \cite{zhu2019modeling}. Roughly 0.5\% of Hawaii's total landmass burns annually, a proportion greater than or roughly equal to any other American state \cite{hawaii2018}. We examine an initial attack fire in the Makaha Valley on the island of Oahu on 19 October 2007, which would go on to burn approximately 500 acres \cite{hnn2007}. The 2007 Makaha Valley wildfire required three days, more than sixty firefighters, and two manned helicopters conducting water drops to contain. Although wildfire perimeter data was not available, the initial wildfire location, historical wind data, elevation map, and fuelbed characteristics were aggregated to simulate and recreate the wildfire. \textit{Fig. 13 shows three photographs of the 19 October 2007 Makaha Valley fire \cite{hnn2007}, along with a satellite photograph of the 10 acre grid where the wildfire began \cite{mapsmakaha2023}, the associated LANDFIRE fuelbed map \cite{landfirefuelbed}, and historical wind directionality trends in the Makaha Valley \cite{makahawinds}. Surface friction typical of mountainous environments such as the Makaha Valley can result in reduced wind speeds.}

\begin{figure}[ht!]
\centering
\includegraphics[width=16.5cm]{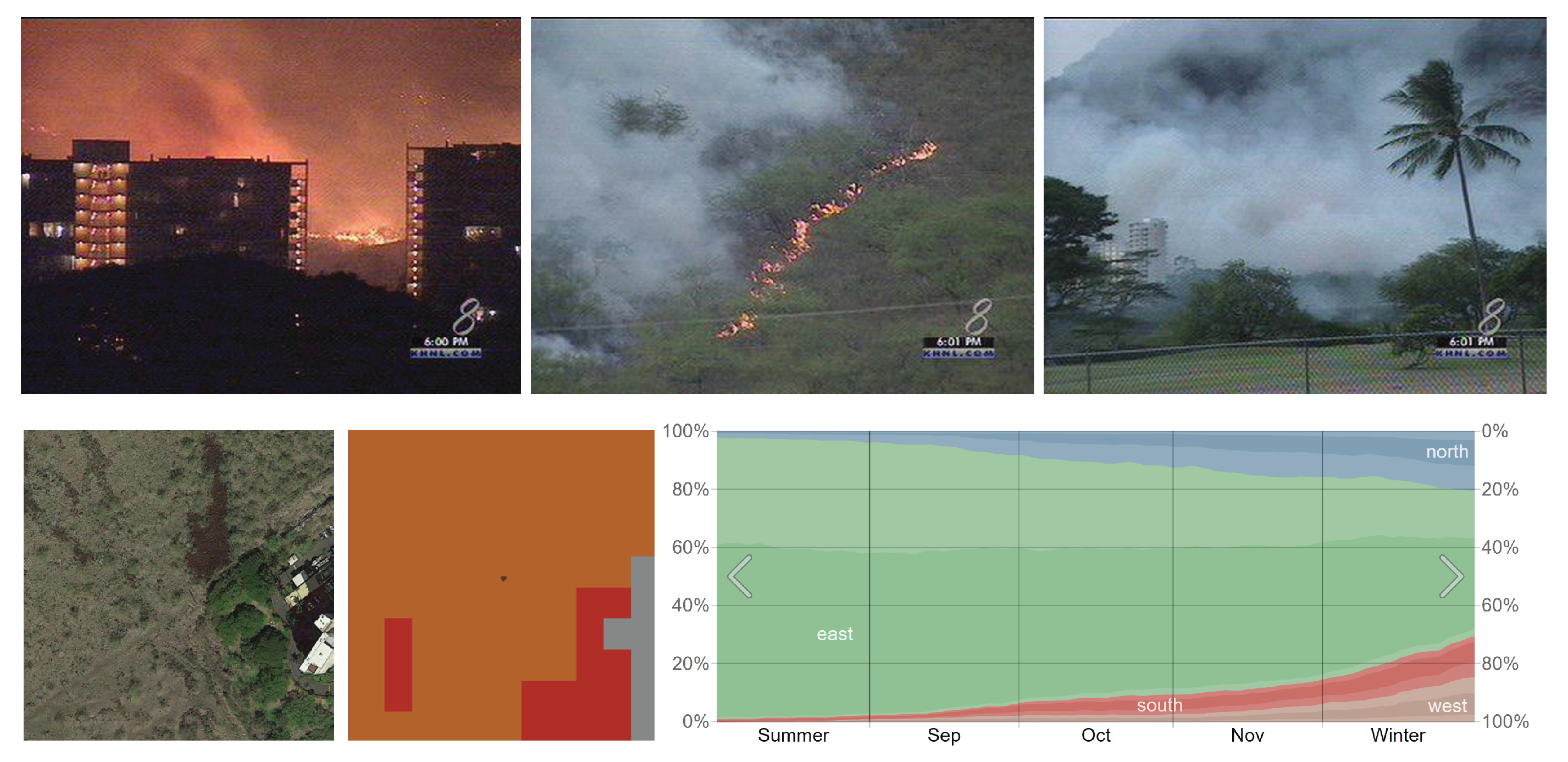}
\caption{Top three: photographs of the 19 October 2007 Makaha Valley fire. From bottom-left to bottom-right: a satellite photograph of the 10 acre grid where the wildfire began, the associated LANDFIRE fuelbed map, and historical wind directionality trends in the Makaha Valley.}
\end{figure}


\subsection{Surveillance}
\noindent
\textbf{Baseline} \\
A myopic baseline is introduced in which unmanned aircraft are jointly rewarded $R_o$ at time-step $t$ for the number of cells observed that had a change in state compared to the belief map. Previous literature typically restricts this reward to only newly identified burning cells, as opposed to newly identified extinguished cells, to encourage surveillance along the wildfire head. However, without an accurate identification of the wildfire tail, suppression assets may be guided to previously extinguished cells. Reward $R_o$ may be weighted to emphasize observations of higher resource cells. Penalties $P_u$, $P_m$, and $P_i$ remain in effect.

\begin{equation}
  R_o = \tau_1 |\{x| B(x) \neq \mathcal{W}(x)\}|
\end{equation}
\begin{equation}
  R_U = R_o - P_u - P_m - P_i
\end{equation}
\\
\noindent
\textbf{Hyperparameters} \\
A three-dimensional grid-search is conducted over a range of discount factors, depths, and exploration constants to identify a high-performing set of three hyperparameters for surveillance MCTS. Table one shows all surveillance hyperparameters and considered values. The model performs well when setting discount factor to 0.95, depth of search to three, and exploration constant to 100. The total iteration limit is 1000 and computation time limit is capped at $\SI{30}{\second}$. The computational time limit must provide sufficient time for the unmanned aircraft to act on the guidance received prior to reevaluation.

\noindent
\textbf{Simulation} \\
Abstracted case study 1 was simulated 20 times with select hyperparameters to compare the myopic baseline against the uncertainty reward model. Suppressive activities were not undertaken to help isolate the effects of both policies without external interference. Two surveillance accuracy metrics were considered: accuracy of the overall belief map relative to the actual wildfire state, and accuracy of the burning cells in the belief map relative to the actual wildfire state.

\begin{table}[h!]
\centering
\caption{Surveillance planner hyperparameters and considered values. Selected values are in bold}
\begin{tabular}{l l}\toprule
 Hyperparameter & Value(s) \\
 \midrule
 computation time limit & \SI{30}{\second} \\
 iteration limit & 1000 \\
 discount factor & \{0.80, 0.85, 0.90, \textbf{0.95}, 0.99\} \\
 depth & \{1, 2, \textbf{3}, 4\} \\
 exploration constant & \{10, 50, \textbf{100}, 200, 1000\} \\
\bottomrule
 \\
\end{tabular}
\end{table}
\subsection{Suppression}
\noindent
\textbf{Baselines} \\
\noindent
Firefighting technique and immediate suppression baselines are introduced for comparison against our approach. Firefighting technique reflects traditional firefighting policy and employs a conditions-based, multi-step approach to suppression. If resources are unevenly distributed, wet-lines are placed around high-value resource areas in an order reflecting their proximity to the wildfire. Thereafter, each drop is aimed at the head of the fire, defined as the distance furthest from the fire origin. The drop selection is made based on increased proximity of the drop to the fire; for example, a north-south drop when placing a wet-line east or west of the fire. Immediate suppression is a myopic policy that rewards actions which maximize the number of wildfire cells extinguished by suppressive activity. MCTS with UCT is applied to optimize immediate suppression policy results.

\noindent
\textbf{Hyperparameters} \\
Three-dimensional grid-searches are conducted to identify a high-performing set of six hyperparameters for suppression MCTS. We first simulate over discount factor, search depth, and exploration constant, then over ASR selection, quantile, and roll-out depth. Table two shows all suppression hyperparameters and considered values. The model performs well when setting search depth to two, exploration constant to 100, ASR selection to two, quantile to 90, and roll-out depth to ten. We note that for a search depth of three outperforms a search depth of two for incipient stages of the wildfire. However, performance for search depth of three falters precipitously as the action space grows with the wildfire. Similarly, during the later stages of the wildfire's growth, a search depth of one outperforms a search depth of two. This suggests an expanding action space may be supported by progressively restricting the search depth or branching factor over the lifetime of the model, through use of double progressive widening or similar \cite{sunberg2018online}. The total iteration limit is 1000 and computation time limit is capped to $\SI{120}{\second}$. The computational time limit must provide sufficient time for the manned aircraft to act on the guidance received prior to reevaluation.

\noindent
\textbf{Simulation} \\
Abstracted case study 1 was simulated 20 times with select hyperparameters to compare the immediate suppression baseline and firefighting technique against our localized and global reward models. Surveillance activities were not undertaken to help isolate the effects of all policies without external interference; perfect information was assumed by equating the belief map to the actual wildfire state. Three suppression metrics were considered: total resources destroyed by the wildfire, wildfire flame size (number of burning grid cells), and average wildfire ring radius. Individual wildfires were then categorized as either fully suppressed, contained, or escaped. A fully suppressed wildfire indicates that suppressive activities were able to extinguish the entirety of the wildfire. A contained wildfire indicates plateauing growth, defined as a wildfire ring size at $t$ = \SI{120}{\minute} within 10\% of the average of the last three (rapid, ultrarapid spread) or four (moderate spread) wildfire ring size states.

\begin{table}[h!]
\centering
\caption{Suppression planner hyperparameters and considered values. Selected values in bold}
\begin{tabular}{l l} \toprule
 Hyperparameter & Value(s) \\
 \midrule
 computation time limit & \SI{120}{\second} \\
 iteration limit & 1000 \\
 discount factor & \{0.80, 0.85, 0.90, \textbf{0.95}, 0.99\} \\
 depth & \{1, \textbf{2}, 3\} \\
  exploration constant & \{10, 50, \textbf{100}, 200, 1000\} \\
 ASR selection & $\{$1, \textbf{2}, 3$\}$ \\
 quantile & $\{$80, \textbf{90}$\}$ \\
 roll-out depth & $\{$5, \textbf{10}, 15$\}$ \\
\bottomrule
\\
\end{tabular}
\end{table}

\subsection{Joint Surveillance and Suppression}
\noindent
\textbf{Simulation} \\
Abstracted case studies 1, 2, and 3 were simulated 20 times with hyperparameters shown in Tables 1 and 2 to compare the immediate suppression baseline and firefighting technique against our localized and global reward models, using imperfection wildfire information attained using the uncertainty surveillance model. Three suppression metrics were again considered: total resources destroyed by the wildfire, wildfire flame size (number of burning grid cells), and average wildfire ring radius.

\subsection{Early Dispatch Procedures}
\noindent
Early dispatch windows are shown as gray boundaries in Fig. 14. Should the predicted wildfire ring size at $t$ = 120 min coincide with this window, a secondary aircraft is dispatched to aid in suppressive activity. The predictive accuracy for a Case 1 model using linear regression is quite good, and stabilizes to within 10\% of the final outcome around 20 minutes when there is no suppressive activity, and within 40 minutes when there is a single aircraft conducting suppression. In situations involving wind shifts, irregular elevation profiles, or uneven fuel maps, a machine-learning model may be integrated to better predict the extent of the wildfire propagation. Fig. 14 also provides insight into the consequence of aircraft suppression across differing wildfire spreads. Unsurprisingly, the less severe the wildfire spread, the greater the impact of added suppressive activity. Perhaps more surprising is the significant impact the addition of a second suppressive aircraft has on all examined wildfires regardless of spread. This suggests a sort of suppression resource ``tipping point'', after which point wildfire containment is likely - assuming optimal use of resources.

\begin{figure}[ht!]
\centering
\includegraphics[width=16.5cm]{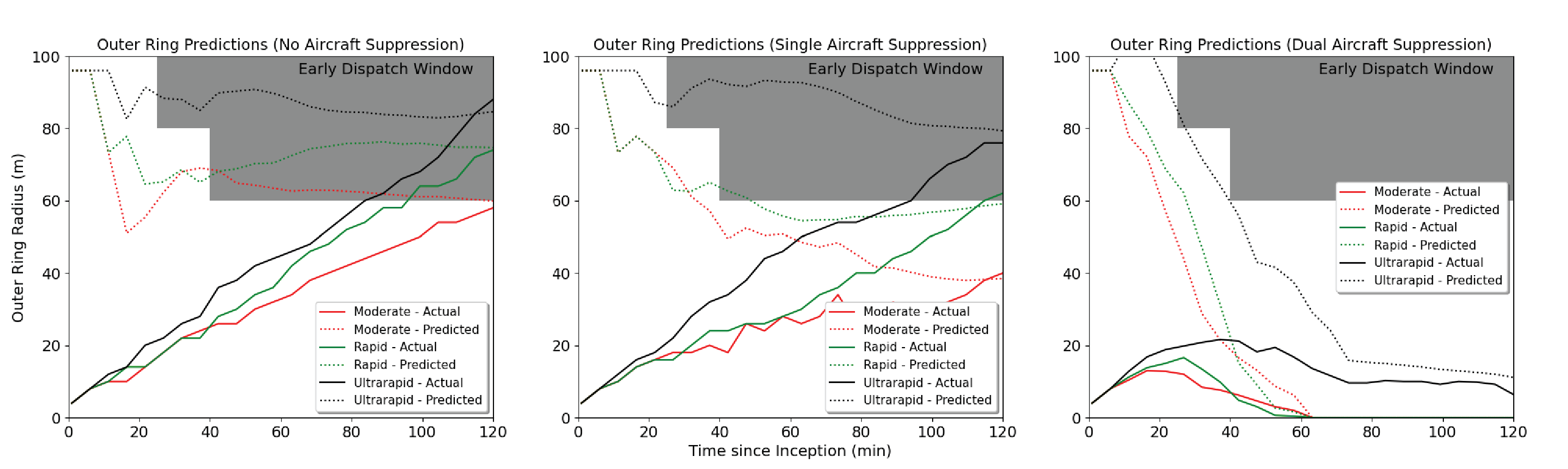}
\caption{Linear regression is used to predict wildfire ring size at $t$ = 120 min for various Case 1 wildfire spread rates with no, single aircraft, and dual aircraft suppression applied.}
\end{figure}

\section{Results}

\subsection{Surveillance}
\noindent
\textbf{Emergent Behaviors} \\
\noindent
Emergence occurs when unique and complex behaviors emerge through the interaction of two or more otherwise simplistic entities. The introduced surveillance planner and associated MCTS solver result in unmanned aircraft exhibiting several emergent behaviors to include dispersion, loitering, circling, and stacking, all shown in Fig. 16. Dispersion occurs when both unmanned aircraft depart from an area of interest to clear room for the manned aircraft, only to return to their original positions immediately following the manned aircraft's departure. Loitering involves one unmanned aircraft avoiding the wildfire entirely while the other maintains full freedom of maneuver around the wildfire. Circling occurs when unmanned aircraft follow one another in a circular pattern. Stacking involves two unmanned aircraft in close lateral proximity with significant altitude between them, such that one unmanned aircraft provides a high-level view of the wildfire while the other provides a condensed low-level view of the wildfire. These emergent behaviors regularly combine with one another (such as circling and stacking), and are typically more prevalent during certain stages in a wildfire's propagation (loitering typically occurs during the initial stages of a wildfire whereas circling and stacking occur during the latter stages). \\
\noindent
\textbf{Simulation Data} \\
As demonstrated in Fig. 15, the uncertainly model increasingly outperforms the belief baseline as the wildfire spread becomes increasingly severe. The belief baseline performs well during early stages of wildfire response (20-40 minutes from inception) and when wildfire spread is slow. When the wildfire is small enough to be local to either unmanned aircraft, querying the immediate vicinity is all that is required, especially with a search depth of up to three. As the wildfire expands beyond the reach of the search tree, weighting the recency of past queries on a by-grid cell basis becomes beneficial. The uncertainty model does this by maintaining surveillance recency data across the entire wildfire in a regularly updated uncertainty map. Given that the uncertainty model rivals or exceeds the performance of the belief baseline in each stage of wildfire propagation and across all spread rates, the uncertainty model will be used exclusively for later joint surveillance and suppression simulations.

\begin{figure}[ht!]
\centering
\includegraphics[width=16.5cm]{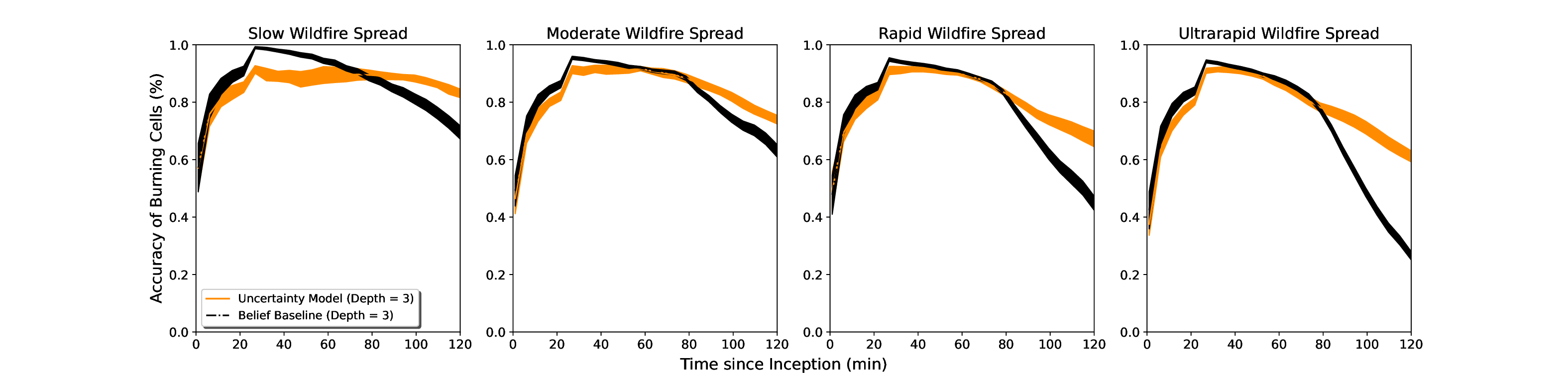}
\caption{Surveillance accuracy of burning cells in \% with 95\% CI ranges for slow, moderate, rapid, and ultrarapid wildfire spread.}
\end{figure}

\begin{figure}[ht!]
\centering
\includegraphics[width=16.5cm]{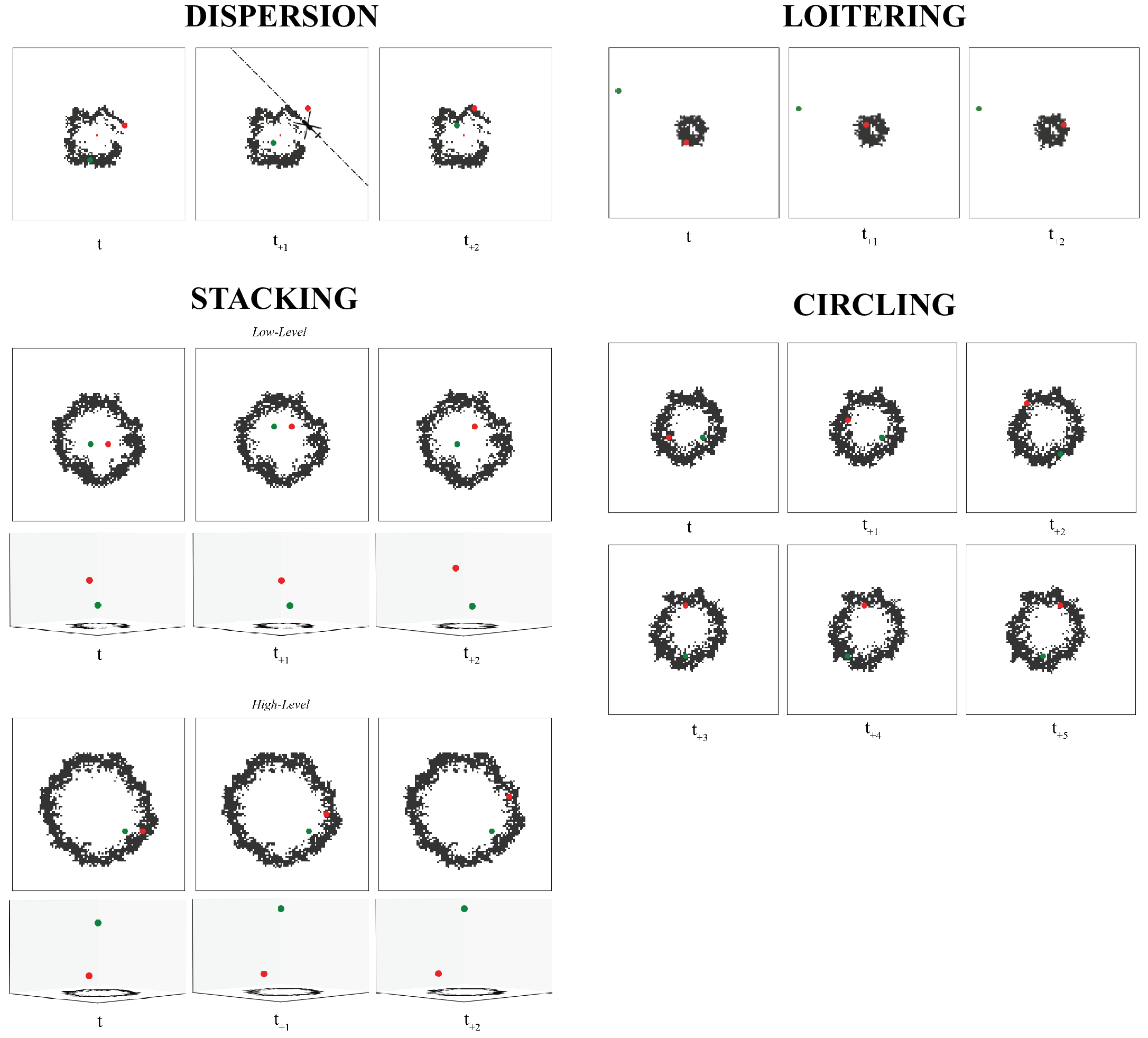}
\caption{Unmanned aircraft exhibiting various forms of emergent behavior to include dispersion, loitering, stacking, and circling.}
\end{figure}

\subsection{Suppression}
\noindent
\textbf{Policy Behavior} \\
The three suppression policy categories under consideration each uniquely prioritize different aspects of the wildfire, as evidenced by their behavior in simulation. Comparison of firefighting technique against optimized destruction minimization policies, shown in Fig. 17, indicates several commonalities however. The destruction minimization policies also place strategic wet-lines, although not with the expediency of the firefighting technique; it prefers to immediately reduce the extent of the initial attack fire rather than preemptively place wet-lines alongside high-value areas. This extinguishes the fire at its infancy, when the greatest effect can be had through suppression. Otherwise, the destruction minimization policies also prioritize the head and exterior of the fire. Destruction minimization policies typically select wet-lines that simultaneously suppress at least some portion of the wildfire, rather than treating wet-lines and suppression as largely exclusive. The immediate suppression baseline is the most myopic of the three categories, as it discards any consideration of future outcomes in favor of extinguishing the greatest amount wildfire in the present.

\begin{figure}[ht!]
\centering
\includegraphics[width=16.5cm]{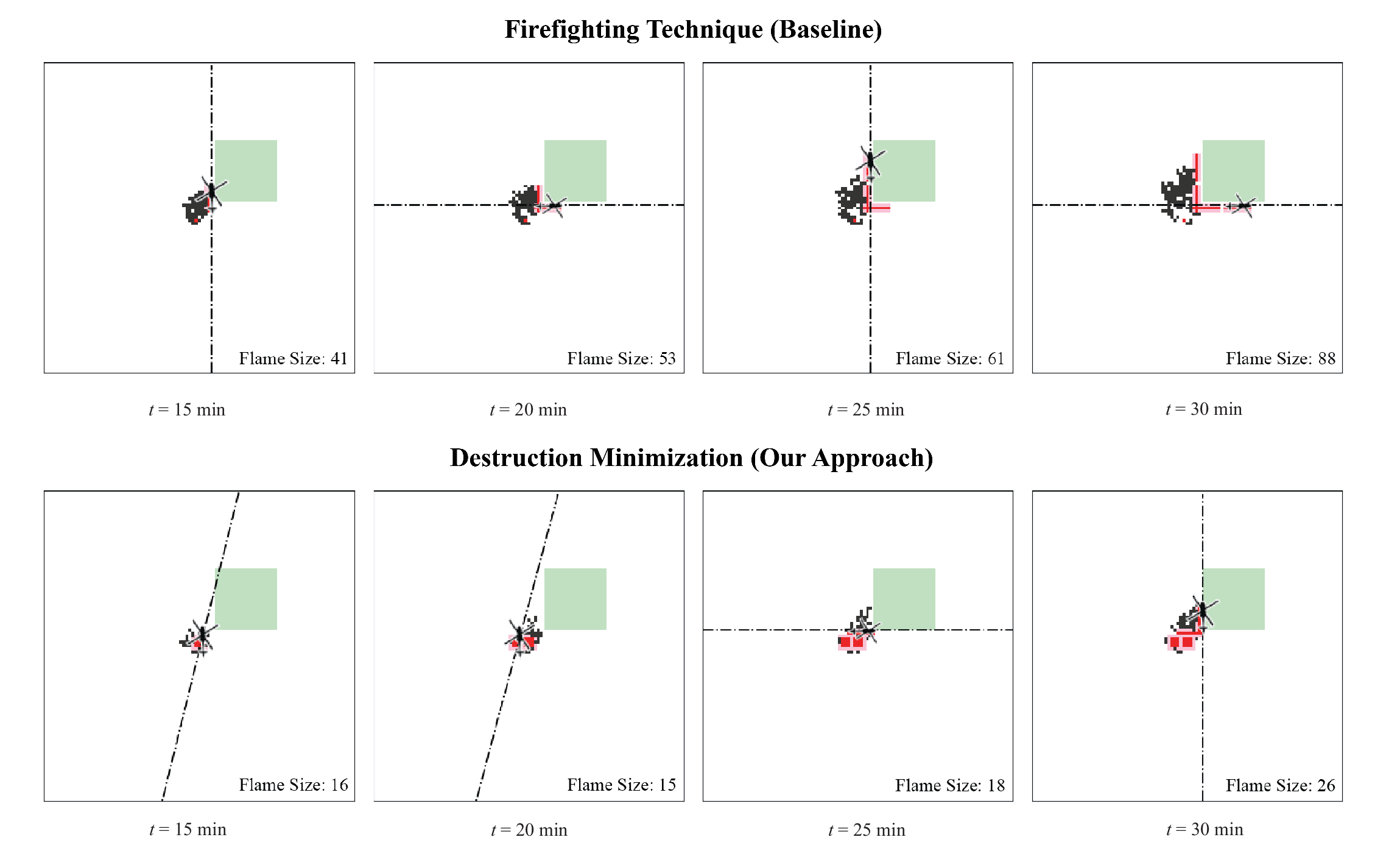}
\caption{Limitations of an overly proactive approach in firefighting technique compared to the strategically-proactive approach in our destruction minimization models.}
\end{figure}

\noindent
\textbf{Simulation Data} \\
As shown in Fig. 18 and 19 and in Table 3, the localized and global resource destruction minimization policies outperform the immediate baseline across all wildfire spreads, and are likely to outperform firefighting techniques in moderate and rapid wildfire spreads. Both forms of resource destruction minimization appear to perform about evenly. Resource destruction minimization suppression results vary increasingly as the wildfire spread becomes increasingly severe; this may be attributed to the unevenness between wildfires that are fully suppressed and those that escape.

\begin{figure}[ht!]
\centering
\includegraphics[width=16.5cm]{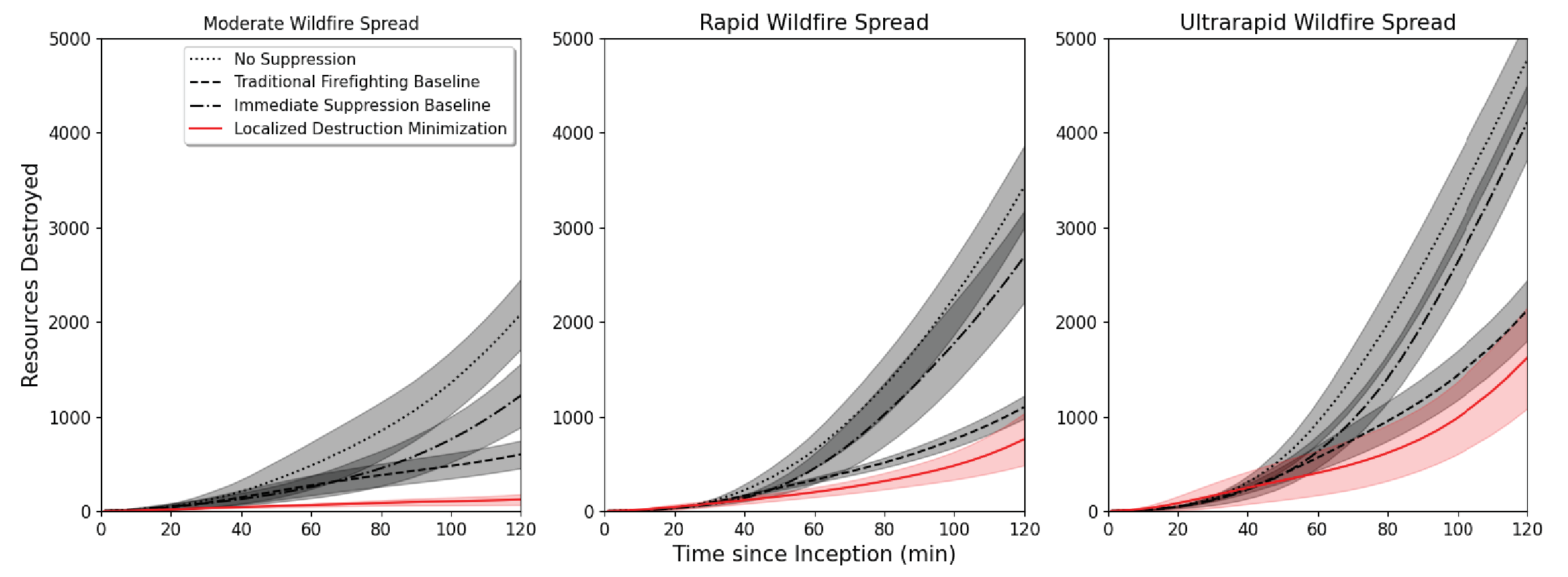}
\caption{A 20-run average comparison of suppression policy selection on resources destroyed over time with 95\% CI ranges for moderate, rapid, and ultrarapid wildfire spreads in Case 1 given a perfect surveillance information assumption.}
\end{figure}

\begin{figure}[ht!]
\centering
\includegraphics[width=16.5cm]{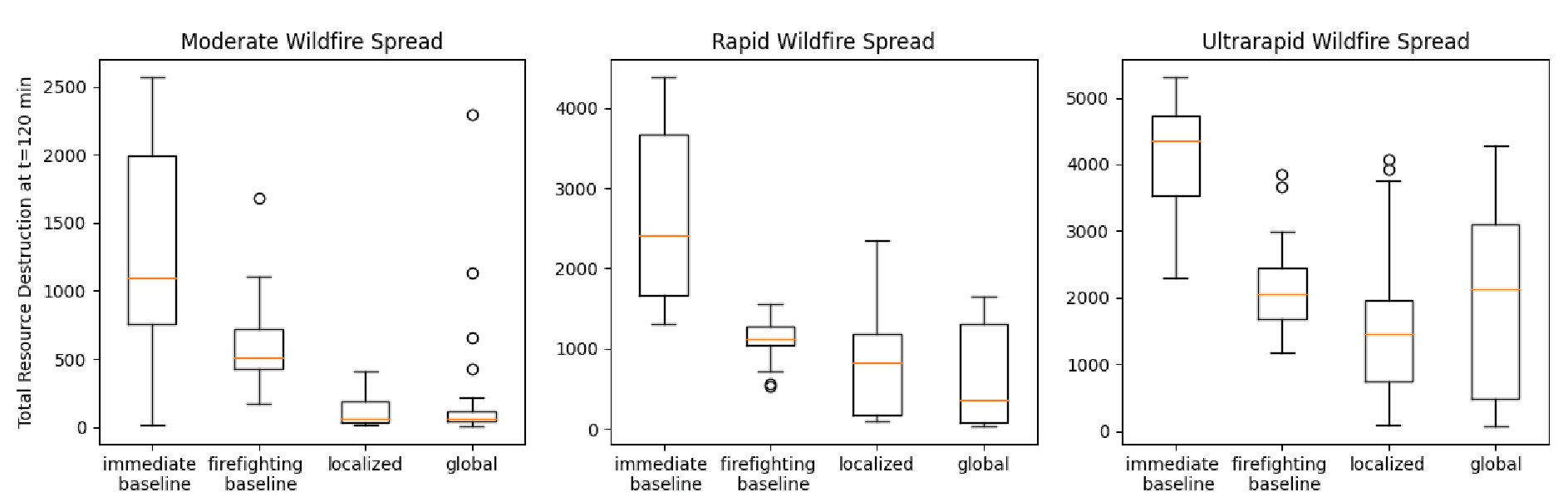}
\caption{A 20-run comparison of suppression policy selection on resources destroyed at $t$=120 min for moderate, rapid, and ultrarapid wildfire spreads in Case 1 with a perfect surveillance information assumption.}
\end{figure}

\begin{table}[h!]
  \begin{center}
    \caption{Suppression Performance (Final Destruction), Case 1, 20 Run Average, 95\% CI}
    \begin{tabular}{l c c c}
    \toprule
      \textbf{Suppression Method} & \textbf{Moderate Fire Spread} & \textbf{Rapid Fire Spread} & \textbf{Ultrarapid Fire Spread} \\
      \textit{Perfect Information Assumption} & & & \\
      \midrule
      \textit{Baselines} & & & \\
      Firefighting Technique       & 600.40 $\pm$ 146.52 & 1100.18 $\pm$ 121.02 & 2120.49 $\pm$ 320.49 \\
      Immediate Suppression & 1223.14 $\pm$ 334.92 & 2687.47 $\pm$ 483.72 & 4107.61 $\pm$ 393.40 \\
      \textit{Our Methods} & & & \\
      Localized Destruction Minimization & \textbf{122.66 $\pm$ 55.57} & \textbf{760.37 $\pm$ 275.75} & 1617.31 $\pm$ 532.71\\
      Global Destruction Minimization & \textbf{272.49 $\pm$ 241.16} & \textbf{669.38 $\pm$ 274.02} & 1910.78 $\pm$ 628.51 \\
      \bottomrule
    \end{tabular}
    \begin{tablenotes}
      \small
      \item Bold indicates methods that outperform both baselines by a statistically significant margin  ($\alpha$ = 0.05)
    \end{tablenotes}
  \end{center}
\end{table}

\subsection{Joint Surveillance and Suppression}
\noindent
\textbf{Simulation Data} \\
Localized and global resource destruction minimization policies outperformed the immediate baseline for moderate wildfire spreads and are likely to outperform the immediate baseline for rapid and ultrarapid wildfire spreads. The localized and global resource destruction minimization policies are also likely to outperform firefighting techniques in moderate wildfire spreads, and are on par for rapid and ultrarapid wildfire spreads. Both forms of resource destruction minimization continue to perform about evenly. Notably, imperfect as opposed to perfect surveillance information has a significantly more detrimental effect on the optimized policies as opposed to firefighting technique. Accurate wildfire data is required to fight the wildfire in a manner that results in full suppression, whereas placing wet-lines along boundaries and high-value areas is more forgiving. These results hold for Cases 2, 3, and 4, despite the increase in environmental complexity. It may be inferred that improving surveillance accuracy is among the most effective ways to optimize suppression results.

Fig. 20 analyzes the end-state of the propagated wildfires post-suppression by bucketing them into one of three categories: fully suppressed, contained, and escaped. Escaped fires represent between 1 and 17\% of all wildfires in the United States, but result in 97\% of the overall burned landmass \cite{reimer2019measuring}, \cite{calkin2005forest}. Escaped fire containment requires a significant expansion in suppression capability by firefighting agencies through the employment of multiple ground and air assets. As shown, the percentage of escaped wildfires in moderate, rapid, and ultrarapid wildfire spreads all greatly exceed 1 to 17\% when firefighting techniques are applied. Firefighting technique resulted in containment in 5 to 15\% of wildfire simulations across all spreads, outperforming the immediate suppression baseline and no suppression whatsoever. This indicates our focus on initial attack fires with the potential to become escaped fires. For reference, the joint surveillance and suppression framework with an optimized resource destruction minimization model applied results in a 100\% full suppression rate when the wildfire spread corresponds to an 83\% full suppression rate using firefighting techniques. Fig. 22 also demonstrates the effectiveness of resource destruction minimization models, with and without early dispatch, relative to the immediate suppression baseline and firefighting techniques. The only instances of fully suppressed wildfires, across all three wildfire spreads, were when resource destruction minimization models were applied. The moderate wildfire spread chart does not feature an early dispatch variation of the destruction minimization model due to the slow propagation sequence not triggering the selected early dispatch window.

\begin{figure}[ht!]
\centering
\includegraphics[width=16.5cm]{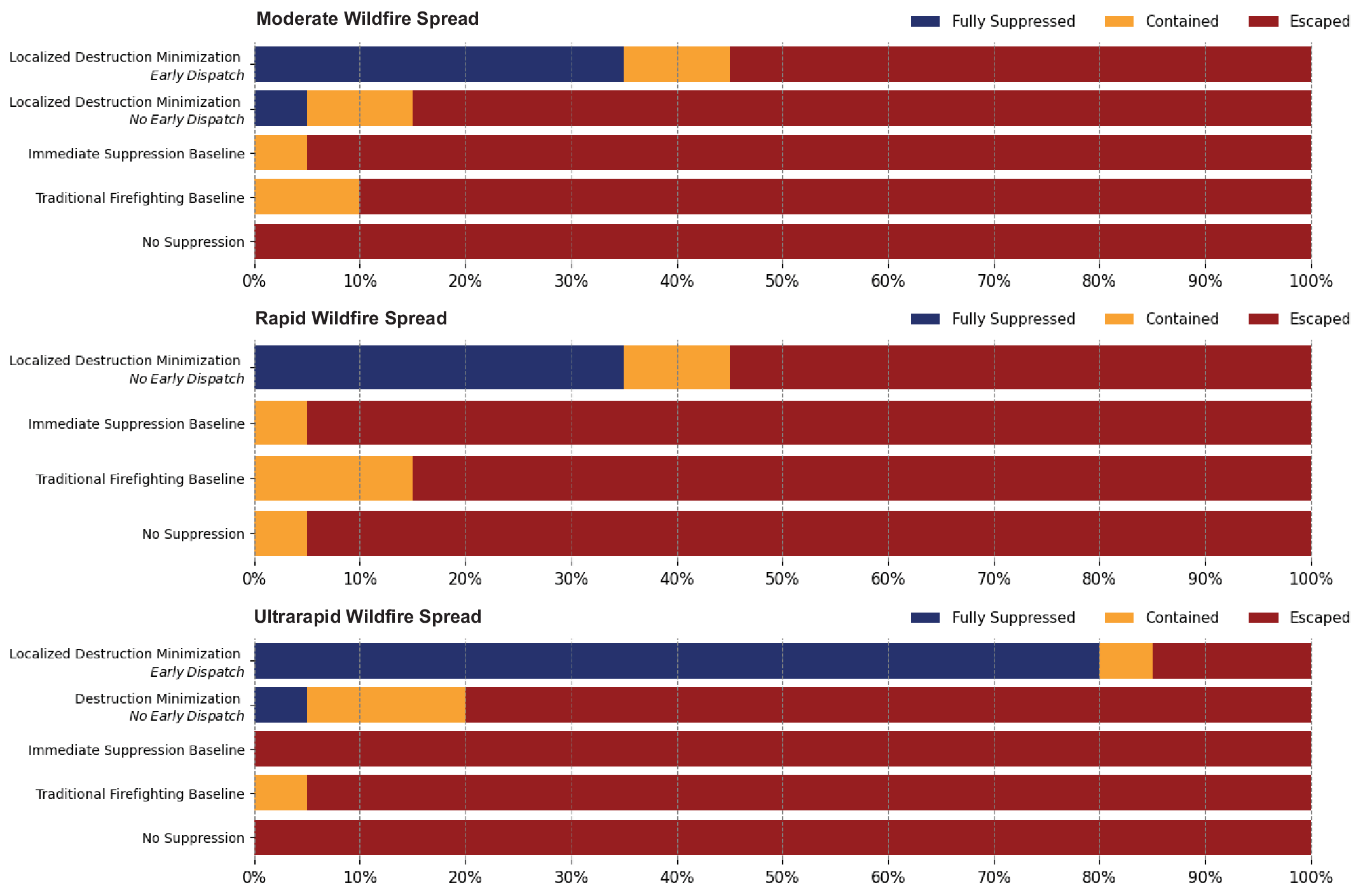}
\caption{A 20-run comparison of suppression policy selection to include early dispatching on wildfire status category at $t$=120 min for moderate, rapid, and ultrarapid wildfire spreads in Case 1 with imperfect surveillance information.}
\end{figure}

\section{Discussion}

\subsection{System Capability}

\noindent
The introduced hierarchical framework is designed to integrate unmanned surveillance aircraft into existing initial attack operations with minimal disruption to participating manned aircraft. It is therefore important to consider the assumptions that enable this capability to function as intended. Those assumptions are related to network architecture, autonomy modes in the case of network degradation, free and stable communication, and enabling hardware.

A MPOMDP, as opposed to a Dec-POMDP, has no impediment to communication \cite{kochenderfer2015decision} - and yet no network is infallible. In our framework, communication exists between unmanned aircraft, between manned aircraft, and between unmanned and manned aircraft. Although there are several feasible approaches to aircraft network architecture, we propose a dual-layer system. The low-level layer is a decentralized mesh network with self-healing properties where aircraft communicate their location in real-time to other networked aircraft. This robust layer aids in collision avoidance. The high-level layer is hierarchical and mirrors the introduced framework. One unmanned aircraft is assigned to be the unmanned network lead. This assignment may rotate such that the unmanned aircraft most central to the mesh, in terms of relative position or strength of network connection, becomes the lead. The unmanned network lead is responsible for: 1.) receiving and fusing observation data across unmanned aircraft, 2.) updating the wildfire belief map, 3.) resolving the surveillance planner using the updated wildfire belief map, 4.) disseminating surveillance guidance to unmanned aircraft, and 5.) disseminating the updated wildfire belief map to all aircraft. A manned network lead may be assigned if there are multiple participating manned aircraft. Alternatively, manned aircraft may independently resolve their own suppression planners. This suits when one manned aircraft is actively suppressing the fire while the other is replenishing water off-station.

Human-autonomous frameworks typically traverse an autonomy hierarchy in response to changes in the network and operating environment. This has significant repercussions for system scalability and robustness. While the ideal case assumes a shared wildfire belief map and centralized decision-making within homogeneous agent groups, this need not be true in our framework. Individual decision processes can be resolved independently if the high-level network layer is degraded. This highlights one advantage of dividing the larger MPOMDP into smaller MDPs. Given lapses in communication with the respective network lead, all aircraft can proceed, if sub-optimally, to the best of their independent ability and per their individual MDPs and belief maps. If the network lead is damaged, another aircraft assumes the network lead role. If a non-lead aircraft is damaged, the framework continues to operate unabated, albeit with fewer observations or suppression actions. Separately, it is critical that the low-level network layer remains intact. An unmanned aircraft that disconnects from the decentralized mesh network, and who is therefore no longer receiving location information, must depart the wildfire vicinity immediately or risk collision. Given the inherent risks involved with manned aircraft integration, we would expect additional fail safe measures to mitigate collision avoidance concerns during communication lapses.

The initial attack is limited to ten acres, or a roughly 200 by 200 meter grid. This is small enough to ensure reliable communication between aircraft using any one of a number of wireless protocols to include WiFi or Bluetooth 5.0. The only exception occurs when the suppression aircraft departs the wildfire to replenish its water bucket. The nearest water replenishing source may be upwards of \SI{10}{km} away, in which case there is limited communication between manned aircraft off-station and the unmanned network lead on-station. This results in the manned aircraft possessing a stale wildfire belief map. Suppression planner performance using the stale wildfire belief map is determined by comparing the frequency of suppression to how quickly the wildfire expands. In slowly and moderately propagating fires, optimizing suppression on a belief map a few minutes old is still likely to provide useful results. Alternatively, the suppression planner can resolve on the condition that the manned aircraft receives an updated wildfire belief map from the unmanned network lead while en-route back to the fire. This ensures a more accurate wildfire belief at the time of suppression planner execution, but reduces the amount of time available for computation. Satellite-based communication or the presence of a nearby ground station can mitigate network reliability concerns.

The initial attack wildfire is constrained in time, and is either suppressed, contained, or becomes an escaped fire at the 120 minute mark. The vast majority of manned suppression aircraft can remain airborne for at least two hours. For example, the S-70 has 150 minutes of flight time when filled with fuel. More limiting is the battery life for unmanned aircraft, especially those with multi-rotors. There exist a handful of higher-end electric and hybrid surveillance quad-copters which have between 55-120 minutes of flight time, depending on the payload. This flight time is expected to incrementally improve with advancements in lithium batteries and composite materials. The presence of recharge stations, or the dispatch a second fleet of unmanned aircraft, can help maintain continuous surveillance operations over the course of an expanding incipient wildfire.

\subsection{Limitations and Future Work}

\noindent
The hierarchical framework introduced, while promising, is subject to certain limitations. These can generally be categorized as model-specific, solver-specific, or domain-specific. From those limitations we identify opportunities for further analysis.

\noindent
\textbf{Model-Specific}

\noindent
MDPs are excellent tools for capturing the dynamics underpinning and uncertainties affecting complex systems, though they are not without limitations. MDPs are subject to the ``curse of dimensionality'', meaning they grow exponentially as the number of states and actions increase \cite{ghavamzadeh2006hierarchical}. Through a combination of wildfire-specific constraints, probabilistic search algorithms, and the decomposition of the MPOMDP into hierarchically arranged sub-problems, we overcame the high-dimensional state and action spaces associated with the initial attack and attained significant and meaningful results. As the initial attack grows beyond the confines of its regulated boundaries and becomes an escaped fire, we would expect the state and suppression action spaces to swell. Additionally, if more aircraft were introduced, surveillance and suppression action spaces would exponentially enlarge. In both cases, the introduced constructs would become less effective. To maintain a similar resolution for surveillance and suppression operations for an escaped fire with an increase in the number of participating aircraft, we suggest an added hierarchical layer, now within homogeneous groupings in addition to just between them. This transforms our hierarchy from an iterating series of planners into an iterating series of sub-hierarchies. We leave a more extensive review of nested MDP hierarchy design and associated computational complexity for the escaped wildfire problem to future research. A second model-specific limitation is surveillance planner frequency and the merging of collision avoidance penalties and wildfire surveillance rewards into one objective function. While the duration between surveillance decisions can be easily modified, there needs to be sufficient time between decisions to 1.) apply the solver and get results, and 2.) actually execute surveillance operations and attain observations. This time scale may differ from that required for robust collision avoidance, which is like to operate at a much higher frequency. In application, the frequency of location sharing between aircraft would be greater than that of observation sharing. We therefore suggest a low-level ``detect and avoid'' system for each unmanned aircraft. The surveillance planner introduced then effectively keeps unmanned aircraft away from the expected manned aircraft axis of advance, while the low-level detect and avoid system informed by location sharing data adjusts for unexpected changes in manned aircraft trajectory.

\noindent
\textbf{Solver-Specific}

\noindent
MCTS is an online planning algorithm and therefore an effective solver choice for the initial attack problem. MCTS can account for non-stationary behavior in the initial attack wildfire. This includes instantaneous changes in wildfire propagation direction and intensity, as demonstrated with the randomized wind-shift simulated in abstracted case study two. Despite its upside, MCTS has certain limitations that must be addressed. MCTS suffers from sample inefficiency in large search spaces, is prone to high-variance, and faces domain-specific challenges related to its internal model. We address state aggregation via nested hierarchical design in the model-specific limitation section above, which can minimize an otherwise expansive search tree, and enable MCTS to obtain accurate statistics and make informed decisions. MCTS roll-outs can vary significantly due to the random nature of simulations. This may result in inconsistent action value estimations and induce problematic noise. Parallelization, variance reduction, and progressive widening techniques can mitigate these concerns. MCTS uses an internal model to conduct simulations, and solver performance is therefore a consequence of how well that internal model reflects reality. This is not an issue when MCTS is used for deterministic games, but becomes concerning when applied to environments rife with uncertainty. The disparity between reality and the internal model is exacerbated with depth. As such, a typical disadvantage of model-based methods is that compounding errors make long-horizon roll-outs unreliable. The depth considered during surveillance planning is shallow enough to avoid these compounding errors, while still remaining useful due to the frequency of reevaluation.

\noindent
\textbf{Domain-Specific}

\noindent
While we have sought to develop a high-fidelity initial attack model, we have not so far addressed fire intensity and flame height, smoke plumes, mixed fleet operations, and the ember attack. Unmanned aircraft attain higher resolution observation data by descending, but doing so may put them at risk of catching fire. Descending below an established safety distance from the ground may result in a penalty applied to the violating aircraft's objective function. A more tailored approach involves modifying the safety distance as a function of flame intensity or height for a given cell column. We have assumed a binary mapping of the wildfire state, but can reasonably include an assessment of both flame intensity given the appropriate sensors, or flame height using three-dimensional wildfire propagation and observation models. \citeauthor{pham2018distributed} apply a safety distance to unmanned aircraft tracking a wildfire, and also introduce a flame intensity model as an artificial potential field \cite{pham2018distributed}. Smoke plumes obscure surveillance aircraft vision-based observations, induce significant errors in sensor measurements \cite{yuan2015survey}, and modify manned suppression aircrafts' axis of advance. Smoke sensors or low-cost cameras outfitted to unmanned aircraft may be used to develop three-dimensional keep-out geofences \cite{kim2022airspace} for smoke plumes. We address agent heterogeneity by separating available assets into homogeneous manned and unmanned aircraft groups. The suppression patterns and partial suppression probabilities in this paper are tailored to a 660 gallon water bucket on a short-line hauled by a S-70 Firehawk helicopter. These patterns and probabilities may be adjusted to any manned aircraft and water suppression platform. We would expect fundamentally different results for a CH-47 Chinook hauling a 2,600 gallon water bucket on a long-line. As the initial attack wildfire escapes, a variety of new manned aircraft arrive on station to include fixed-wing tankers, helicopters carrying various sizes of water buckets, and smoke-jumper aircraft. The set of joint actions across the diversity of manned aircraft may give rise to interesting emergent behaviors, but must also comply with Fire Traffic Area altitudes, orbits, and routing structures. The set of considered neighboring cells during wildfire propagation may be expanded to incorporate the spread of embers across sizable distances in the case of an ember attack. An ember attack occurs when wildfires in strong wind conditions carry embers beyond the fire front. We focus on the general form of the initial attack and therefore consider only adjacent neighbors during propagation. MDP models exist that have incorporated airborne ember spread for larger wildfires \cite{julian2019distributed}.

\section{Conclusion}
\noindent
The coordination of wildfire surveillance and suppression activities using manned and unmanned aircraft in tandem is an effective means of reducing wildfire propagation severity and minimizing wildfire destruction. A hierarchical framework involving iterating surveillance and suppression planners is introduced to divide an otherwise intractable MPOMDP into optimizable sub-problems acting on asynchronous but otherwise consistent time scales. Surveillance, suppression, and joint surveillance and suppression models with unique MCTS-extensions applied are compared in simulation across abstracted and actual case studies. We demonstrate how a hierarchical approach to Markov decision processes may be used to ensure collision avoidance between unmanned and manned aircraft operating in close proximity, and how teaming aerospace operations may extend into wildfire management. We further find that our hierarchical framework with a resource destruction minimization reward model applied significantly outperforms firefighting techniques and a myopic baseline in preventing initial attack fires from developing into escaped fires.

\section*{Funding Sources}
\noindent
No sponsorship or financial support was provided in the development of this manuscript.

\section*{Acknowledgments}
\noindent
We thank CALFIRE and the C/3-25 Aviation Regiment at Wheeler Army Airfield, Hawaii for their collective insight into initial attack aerial firefighting operations.

\newpage

\newpage

\section*{Additional Materials}

\begin{figure}[ht!]
\centering
\includegraphics[width=13.8cm]{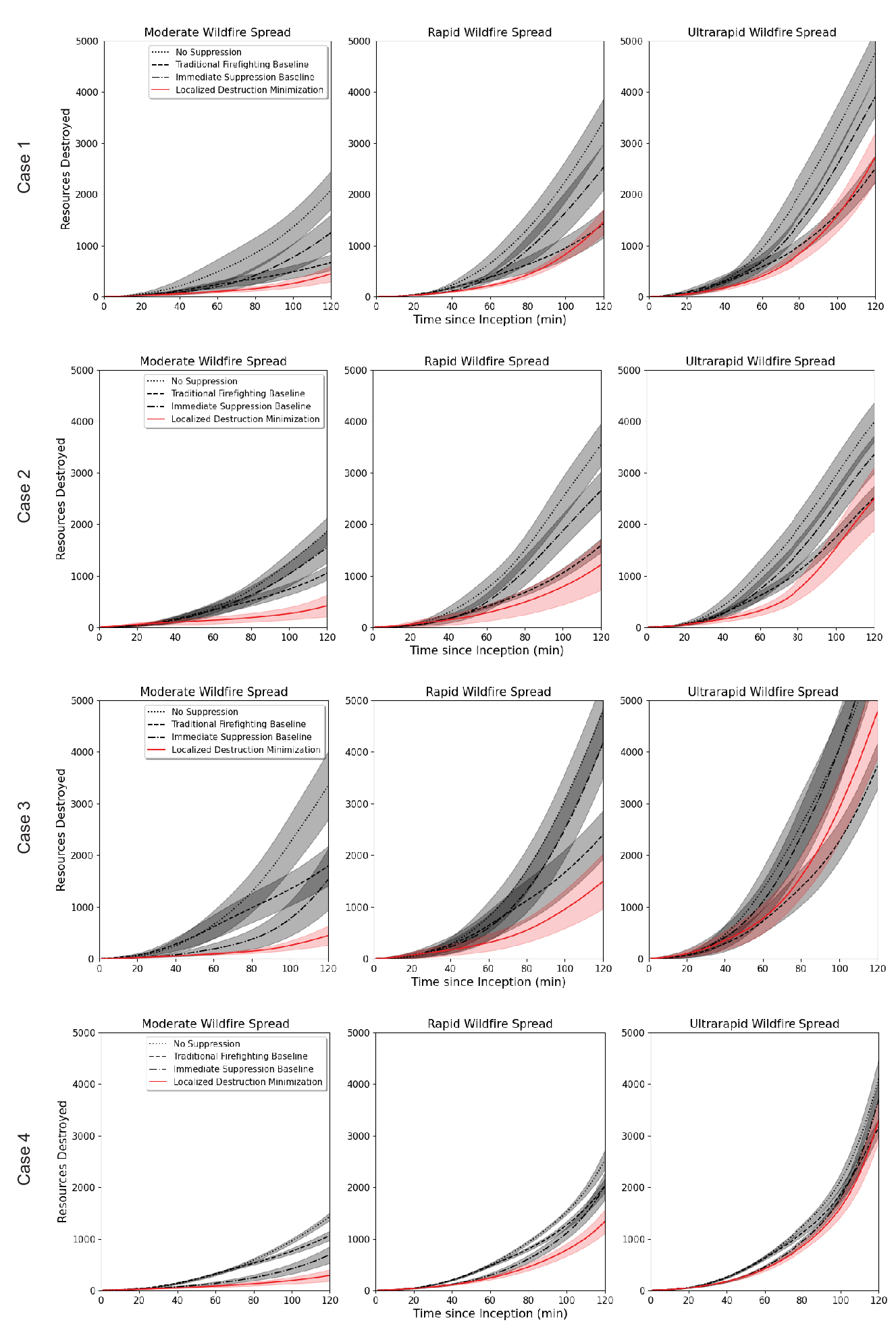}
\caption{A 20-run average comparison of suppression policy selection on resources destroyed over time  with 95\% CI ranges for moderate, rapid, and ultrarapid wildfire spreads in all four cases with imperfect surveillance information.}
\end{figure}

\begin{figure}[ht!]
\centering
\includegraphics[width=16cm]{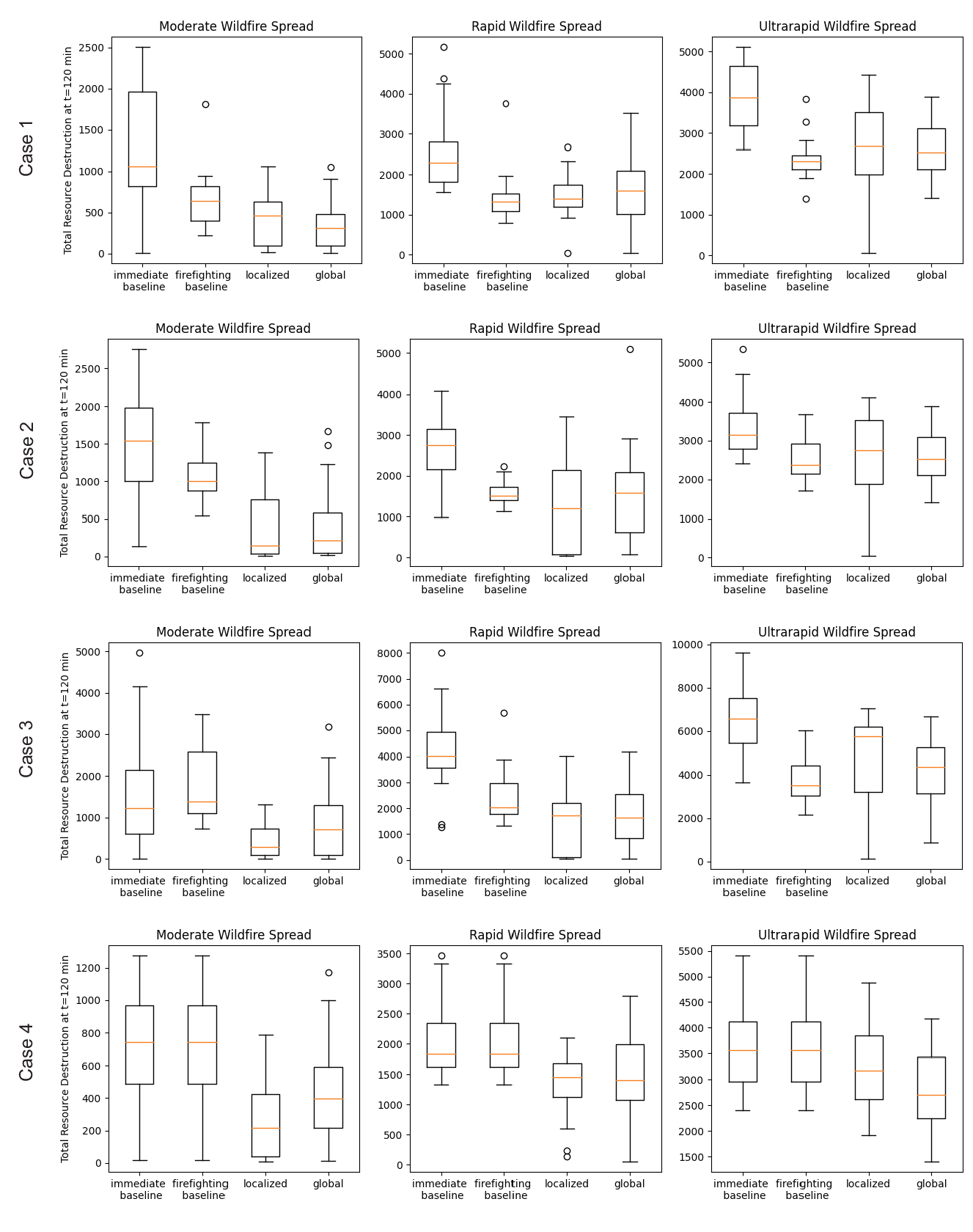}
\caption{A 20-run comparison of suppression policy selection on resources destroyed $t$=120 min for moderate, rapid, and ultrarapid wildfire spreads in all four cases with imperfect surveillance information.}
\end{figure}

\begin{table}[h!]

  \begin{center}
    \caption{Joint Surveillance-Suppression Performance (Final Destruction), 20 Run Average, 95\% CI}
    \begin{tabular}{S l c c c}
    \hline \hline
      \textbf{Case} & \textbf{Suppression Method} & \textbf{Moderate} & \textbf{Rapid} & \textbf{Ultrarapid}\\
      & \textit{Uncertainty Surveillance Model} & \text{Fire Spread} & \text{Fire Spread} & \text{Fire Spread} \\

      & \textit{Baselines} & & & \\
      & Firefighting Technique       & 663.16$\pm$ 153.86 & 1416.30 $\pm$ 270.56 & 2504.41 $\pm$ 259.58 \\
      1 & Immediate Suppression & 1247.83 $\pm$ 351.86 & 2523.71 $\pm$ 442.56 & 3909.05 $\pm$ 373.87 \\
       & \textit{Our Methods} & & & \\
      & Localized Destruction Minimization & 444.48 $\pm$ 151.70 & 1455.15 $\pm$ 244.14 & 2727.90 $\pm$ 480.10 \\
      & Global Destruction Minimization & \textbf{348.12 $\pm$ 133.91} & 1602.76 $\pm$ 390.03 & 2581.25 $\pm$ 305.61 \\
      \\
      & \textit{Baselines} & & & \\
      & Firefighting Technique       & 1050.80 $\pm$ 134.23 & 1586.44 $\pm$ 130.94 & 2515.49 $\pm$ 227.91 \\
      2 & Immediate Suppression & 1552.63 $\pm$ 298.54 & 2653.19 $\pm$ 364.15 & 3354.86 $\pm$ 364.23 \\
      & \textit{Our Methods} & & & \\
      & Localized Destruction Minimization & \textbf{417.43 $\pm$ 209.84} & 1210.73 $\pm$ 489.17 & 2494.31 $\pm$ 610.34 \\
      & Global Destruction Minimization & \textbf{445.21 $\pm$ 233.31} & 1549.57 $\pm$ 535.60 & 2581.25 $\pm$ 305.61 \\
      \\
      & \textit{Baselines} & & & \\
      & Firefighting Technique       & 1792.59 $\pm$ 382.81 & 2396.35 $\pm$ 465.55 & 3728.30 $\pm$ 441.38 \\
      3 & Immediate Suppression & 1532.93 $\pm$ 592.88 & 4173.03 $\pm$ 677.81 & 6483.51 $\pm$ 665.34  \\
       & \textit{Our Methods} & & & \\
      & Localized Destruction Minimization & \textbf{447.74 $\pm$ 185.17} & \textbf{1493.11 $\pm$ 532.93} & 4778.15 $\pm$ 892.61 \\
      & Global Destruction Minimization & \textbf{934.72 $\pm$ 406.79} & 1753.15 $\pm$ 551.94 & 4145.07 $\pm$ 651.05 \\
      \\
      & \textit{Baselines} & & & \\
      & Firefighting Technique        & 1061.27 $\pm$ 90.40 & 2020.31 $\pm$ 131.77 & 3160.39 $\pm$ 165.56 \\
      4 & Immediate Suppression  & 690.67 $\pm$ 160.69 & 2010.83 $\pm$ 262.97 & 3690.32 $\pm$ 372.43 \\
       & \textit{Our Methods} & & & \\
      & Localized Destruction Minimization & \textbf{288.61 $\pm$ 113.47} & \textbf{1329.51 $\pm$ 226.86} & 3290.26 $\pm$ 407.79 \\
      & Global Destruction Minimization & \textbf{434.65 $\pm$ 143.71} & 1477.92 $\pm$ 321.53 & 2800.06 $\pm$ 332.47 \\
      \hline
      \hline
    \end{tabular}
    \begin{tablenotes}
      \small
      \item Bold indicates methods that outperform both baselines by a statistically significant margin ($\alpha$ = 0.05)
    \end{tablenotes}
  \end{center}
\end{table}

\end{document}